\documentclass[useAMS,usenatbib]{mn2e}

\usepackage[psamsfonts]{amssymb}
\usepackage[dvips]{graphicx}
\usepackage{graphicx}
\usepackage{amsmath,alltt}                                                                                                                          
\usepackage{multirow}
\usepackage{rotating}
\usepackage{lscape}
\usepackage{tablefootnote}
\usepackage{hyperref}
\makeatletter
\newcommand{\@todonotes@enable}{1}
\newcommand{\@todonotes@inline}{1}
\makeatother
\input{notes.tex}

\title[The Effect of Dark Matter Substructure on Clusters]{Modelling the Effects of Dark Matter Substructure on Globular Cluster Evolution with the Tidal Approximation}
\author[Webb et al.]{Jeremy J. Webb$^1$\thanks{E-mail: webb@astro.utoronto.ca (JW)}, Jo Bovy$^1$, Raymond G. Carlberg$^1$ \& Mark Gieles$^{2,3,4}$ \\
$^1$ Department of Astronomy and Astrophysics, University of Toronto, 50 St. George Street, Toronto ON M5S 3H4, Canada \\
$^2$ Department of Physics, University of Surrey, Guildford GU2 7XH, UK \\
$^3$ Institut de Ci\`{e}ncies del Cosmos (ICCUB), Universitat de Barcelona, Mart\'{i} i Franqu\`{e}s 1, E08028 Barcelona, Spain\\
$^4$ ICREA, Pg. Lluis Companys 23, 08010 Barcelona, Spain}
\begin{document}

\pagerange{\pageref{firstpage}--\pageref{lastpage}} \pubyear{2019}

\maketitle

\label{firstpage}

\begin{abstract}
We present direct $N$-body simulations of tidally filling 30,000 ${\rm M}_\odot$ star clusters orbiting between 10 kpc and 100 kpc in galaxies with a range of dark matter substructure properties. The time-dependent tidal force is determined based on the combined tidal tensor of the galaxy's smooth and clumpy dark matter components, the latter of which causes fluctuations in the tidal field that can heat clusters. The strength and duration of these fluctuations are sensitive to the local dark matter density, substructure fraction, sub-halo mass function, and the sub-halo mass-size relation. Based on the cold dark matter framework, we initially assume sub-halos are Hernquist spheres following a power-law mass function between $10^5$ and $10^{11} {\rm M}_\odot$ and find that tidal fluctuations are too weak and too short to affect star cluster evolution. Treating sub-halos as point masses, to explore how denser sub-halos affect clusters, we find that only sub-halos with masses greater than $10^{6} {\rm M}_\odot$ will cause cluster dissolution times to decrease. These interactions can also decrease the size of a cluster while increasing  the velocity dispersion and tangential anisotropy in the outer regions via tidal heating. Hence increased fluctuations in the tidal tensor, especially fluctuations that are due to low-mass halos, do not necessarily translate into mass loss. We further conclude that the tidal approximation can be used to model cluster evolution in the tidal fields of cosmological simulations with a minimum cold dark matter sub-halo mass of $10^{6} {\rm M}_\odot$, as the effect of lower-mass sub-halos on star clusters is negligible.
\end{abstract}


\section{Introduction} \label{intro}

The Cold Dark Matter (CDM) framework predicts that galaxies form hierarchically, with a central dark matter halo assembled via the merging of a large number of dark matter sub-halos \citep{white78, white91, springel05}. Cosmological simulations suggest that while a large fraction of dark matter sub-halos will completely disrupt within a Hubble time due to dynamical friction, tidal heating from the central host halo, and encounters with other sub halos, a non-negligible amount of sub-halos should still exist in the low-density outskirts of galaxies today \citep{ghigna98, moore99, klypin99, diemand08, stadel09}. While the more massive of these sub-halos take the form of luminous satellite galaxies, the majority of the predicted sub-halo population remains unaccounted for as sub-halos that do not contain luminous matter and can only be studied indirectly via gravitational interactions.

Large scale structure simulations like Via Lactea II \citep{diemand08} and The Aquarius Project \citep{springel08} suggest that sub-halos can even be found in the inner regions of a dark matter halo, with the present day substructure fraction of dark matter in Milky Way-like galaxies varying with galactocentric distance between $0.01\%$ within 10 kpc to $10\%$ beyond 100 kpc for sub-halos with masses between $10^5$ and $10^9 {\rm M}_\odot$. However these estimates have been debated, with arguments for both lower and higher substructure fractions being made in the literature. Recent results from the FIRE project \citep{garrisonkimmel17} suggest that the substructure fraction may in fact be \textit{lower}, as baryonic processes may contribute to the disruption of sub-halos. Furthermore, recent work by \citet{kelley18} suggests that the Galactic disk may be responsible for the disruption of a large fraction of sub-halos. Conversely, an in-depth study by \citet{vandenbosch18a} indicates the substructure fraction may instead be \textit{higher}, as the seemingly high disruption rate of sub-halos in cosmological simulations is artificial, an artifact of inadequate force softening leading to high mass loss rates and artificial tidal stripping. \citet{vandenbosch18b} demonstrated that with sufficient force and mass resolution, most sub-halos will survive a Hubble time and that most cosmological simulations will suffer from an overmerging of sub-halos. The estimated distribution of sub-halos in galaxies is also dependent on the nature of dark matter itself, with several alternative dark matter models including warm dark matter, self-interacting dark matter, and 'fuzzy' dark matter still being pursued \citep[e.g.][]{press90, hu00, spergel00, vogelsberger12, elbert15, ludlow16, hui17}. Hence constraining the properties of dark matter substructure is a very important step towards understanding the nature of dark matter as well as the formation and evolution of galaxies. 

The search for observational evidence of dark matter substructure in galaxies is on-going, as traditional methods for in-directly detecting dark matter sub-halos (e.g. modelling tidal streams and gravitational lensing) have yet to agree on either the Milky Way's or an external galaxies' current substructure composition. Gravitational lensing allows for constraints to be placed on the dark matter substructure content of external galaxies, as orbiting sub-halos will result in anomalies in strong gravitational lenses \citep{mao98,dalal02,vegetti12}. The disruption of stellar streams by dark matter sub-halos in the Milky Way has also been well studied \citep[e.g.][]{johnston02, ibata02, carlberg09, erkal16,sanders16,bovy17,carlberg17}, with gaps in stellar streams believed to be a tell-tale sign that a stream has recently encountered a sub-halo. Hence the presence of gaps in a tidal stream, or lack thereof, can be used to constrain the dark-matter substructure properties of the Milky Way \citep{yoon11, carlberg12, erkal15a, erkal15b, bovy16, carlberg16,banik18a, bonaca18}. However gaps in tidal streams, as well as over-densities and asymmetry, can also be produced as stars are tidally stripped from a star cluster along its orbit \citep{kupper10}, disk shocking \citep{odenkirchen03}, spiral arms \citep{dehnen04}, a tri-axial halo \citep{kupper15}, the Galactic bar \citep{pearson17}, interactions with giant molecular clouds \citep{amorisco16}, the stream's progenitor cluster \citep{webb18}, and complex dynamical histories (i.e a time dependent tidal field due to galaxy growth via mergers or accretion) \citep{carlberg18}. \citet{banik18b} recently studied these effects in detail for the Pal 5 stream and found that the bar and molecular clouds can each individually explain the observed structure of the Pal 5 stream. Hence new methods are required in order to help search for dark matter substructure in the Milky Way and constrain its properties.

Globular clusters present a potentially new opportunity in the search for dark matter substructure, as their long-term evolution has long been known to be strongly linked to their host galaxy \citep[e.g.][]{searle78, baumgardt03, prieto08, kruijssen09, rieder13, li18, kruijssen18}. Several studies have shown globular clusters to be susceptible to tidal heating over short \citep{spitzer58, spitzer87, gnedin97, gieles06, lamers06, kruijssen11, webb14b, gieles16, webb18b} and long timescales \citep{baumgardt03, kruijssen09, webb13, webb14a}, both of which would be side effects of globular clusters interacting with dark matter sub-halos.  A recent analytic study by \citet{penarrubia18} specifically found that star clusters will be susceptible to fluctuations in the gravitational field of their host galaxy due to dark matter substructure, with repeated interactions leading to higher cluster mass loss rates. Furthermore, since sub-halo encounter rates increase as sub-halo mass decreases, it is possible that the low-mass end of the sub-halo mass function will contribute the most to the disruption of stellar clusters. However \citet{penarrubia19} predicts that the amount of tidal heating caused by low-mass sub-halo interactions is likely minimal.

Probing the low-mass end of the sub-halo mass function with simulations has, for the most part, remained elusive as most studies on interactions between stellar streams and sub-halos focus on higher-mass sub-halos ($\rm{M} \geq 10^7 {\rm M}_\odot$) \citep[e.g.][]{erkal15a,sanders16,bovy17,carlberg17}

Constraining how GCs are affected by dark matter substructure is also important when modelling the evolution of clusters in cosmologically motivated tidal fields. Large-scale simulations of galaxy formation are now capable of resolving the formation sites and orbital evolution of GCs \citep[e.g.][]{kravtsov05, maxwell12, renaud17, li17, pfeffer18, kim18, mandelker18}, which allows for the tidal field at the cluster's location to be tracked and the small-scale evolution of individual clusters to be modelled afterwards. Knowing how GCs are affected by sub-halos will set the necessary mass, spatial, and time resolution that a galaxy simulation must have in order to use its tidal field to accurately model the evolution of a GC.

In this study we explore how dark matter sub-halos affect GC evolution in order to bridge the gap between large-scale models of galaxy formation and the external tidal fields used in small-scale models of GC evolution, potentially opening the door for observations of GCs to be used as new tools to search for evidence of dark matter substructure. In Section \ref{s_method} we describe our method for modelling star clusters in galaxy potentials containing substructure. In Sections \ref{s_results} we explore how interactions with dark matter sub-halos affect the evolution of cluster mass, size, velocity dispersion and orbital anisotropy as a function substructure fraction. We also investigate how sensitive our findings are to the sub-halo mass function and the structural properties of individual sub-halos. In Section \ref{s_discussion} we quantify how the presence of dark matter substructure affects the tidal tensor experienced by GCs and the subsequent effect the modified tidal tensor has on cluster evolution. Finally, we summarize our findings in Section \ref{s_conclusion}. 

\section{Method} \label{s_method}
\subsection{The Tidal Approximation}

To determine how interactions with dark matter sub-halos will alter the evolution of a GC, we perform simulations of clusters orbiting in galaxies containing sub-halos with a range of initial properties. More specifically we vary the dark matter substructure fraction, the range of allowable sub-halo masses and the structural properties of the sub-halos. To simulate the star clusters, we make use of the direct $N$-body code NBODY6tt \citep{renaud11}, a modified version of NBODY6 \citep{aarseth03} which allows for a tidal tensor to be used to model the external tidal field experienced by the cluster. In the tidal approximation, a star's position $\rm r^{\prime}$ relative to the centre of its host star cluster is assumed to be much smaller than the scale over which the external tidal field varies. This condition allows for the star's acceleration in a cluster-centred reference frame due to an external potential $\Psi$ to be written as:

\begin{equation}
\frac{\rm{d}^2 \mathbf{r^{\prime}}}{\rm{d}t^2} = -\nabla \Psi(\mathbf{r^{\prime}})+\nabla \Psi(0) = \mathbf{T}_{\rm t}^{ij}(\mathbf{r^{\prime}}) \cdot \mathbf{r^{\prime}}
\end{equation}

where $\mathbf{T}_{\rm t}^{ij}(\rm r^{\prime})$ is the tidal tensor, which has the more explicit form $\mathbf{T}_{\rm {t}}^{ij}(\rm r^{\prime})= -\frac{\rm{d}^2\Psi(r^{\prime})}{\rm{d}q^i \rm{d}q^j}$ (e.g. in Cartesian coordinates $\rm{q}^1$, $\rm{q}^2$ and $\rm{q}^3$ will equal  x,y, and z). Hence knowing the time evolution of the tidal tensor allows for the force acting on each star in a cluster, relative to the force experienced by the cluster's centre of mass, to be determined at all times. To model the effects of dark matter substructure on cluster evolution, we first need to generate galaxy models consisting of a smooth and clumpy dark matter component and be able to calculate the tidal tensor anywhere in the galaxy. To this end, a new feature has been added to the publicly available code \texttt{galpy} \footnote{http://github.com/jobovy/galpy} \citep{bovy15} that allows for the tidal tensor to be calculated at any given location within a potential. Hence it is possible to simply place a star cluster within a model galaxy, integrate its orbit using \texttt{galpy}, and compute the time evolution of the tidal tensor at the cluster's centre. In the following sections, we will describe both the galaxy models and the star cluster simulations used in this study.

\subsection{Galaxy Model}\label{sec:gal}

In all cases, the host galaxy is taken to be logarithmic halo potential of the form:

\begin{equation}
\Psi(\mathrm{r}_{\rm gc})=\rm v_c^2 \ln(r_{\rm gc})
\end{equation}

where $\rm r_{gc}$ is galactocentric distance and $\rm v_c = 220$ km/s to mimic the Milky Way \citep{bovy12}. The potential is then broken down into a smooth component ($\Psi_{\rm smooth}(\mathrm{r}_{\rm gc})$) and a component due to substructure ($\Psi_{\rm sub}(\mathrm{r}_{\rm gc})$). For a star cluster at $\rm r_gc$ and a given sub-structure fraction $\rm{f}_{\mathrm{\rm sub}}$, the contribution to each element of the tidal tensor $\mathbf{T}_{\rm{t}}^{ij}(\mathrm{r}_{\rm gc})$ from the smooth component of the galaxy will simply be $\mathbf{T}^{ij}_{\rm{t},\mathrm{\rm smooth}}(\mathrm{r}_{\rm gc})=-(1-\rm{f}_{\mathrm{\rm sub}}) \frac{\rm{d}^2\Psi(r_{\rm gc})}{\rm{dq}^i \rm{dq}^j}$. 

To determine the contribution from dark matter sub-structure, we need to populate the galaxy model with dark matter sub-halos. Assuming a power-law mass function of slope $-1.9$ \citep{springel08}, the number of sub-halos in our galaxy model will depend directly on our choice of the lower sub-halo mass limit, which CDM predicts could go down to planet like masses. Unfortunately, such a wide mass range would result in a sub-halo population that is too computationally expensive to evolve. Instead, we have elected to only include sub-halos with masses between $10^5$ and $10^{11} {\rm M}_\odot$ (the typical mass range used in cosmological simulations) which for substructure fractions between $1\%$ and $10\%$ yields at most a few hundred thousand sub-halos for which their orbits can be evolved for 10 Gyr in a reasonable amount of time. We explore how lower-mass sub-halos affect the evolution of our model clusters by analysing how each mass range contributes to the degree of fluctuation in each tidal tensor. As we will show in Section \ref{sec:mass_function}, sub-halos with masses between $10^5$ and $10^{6} {\rm M}_\odot$ contribute very little to the mass loss rate experienced by a cluster, which suggests it is not necessary to probe below the adopted lower sub-halo mass limit of $10^5 {\rm M}_\odot$ when modelling the effects of cold dark matter substructure on cluster evolution.

First, we generate a position for each sub-halo in the galaxy such that the combined mass profile of smooth dark matter and dark matter substructure follows the exact mass profile expected for an isothermal potential out to 200 kpc. Individual sub-halo velocities are drawn from a Gaussian distribution with a dispersion of 220/$\sqrt{2}$ km/s. Motivated by the properties of sub-halos in cosmological simulations \citep{diemand08, springel08}, we first assume that dark matter sub-halos can be modelled as Hernquist spheres that follow a mass-size relation of the form $\rm{r_s}=1.05$ kpc $(\rm{M}/10^8 {\rm M}_\odot)^{\frac{1}{2}}$ \citep{bovy17}, where $\rm{r_s}$ is the scale radius of the Hernquist sphere. Having $\rm{r_s} \propto \rm{M}^{1/2}$ ensures that dark sub-halos have a constant surface density, regardless of their mass, consistent with \citet{donato09}. The substructure contribution to $\mathbf{T}_{\rm t}^{ij}(\mathrm{r}_{\rm gc})$ will then simply be the sum of the tidal tensors produced by each sub-halo $\mathbf{T}^{ij}_{\mathrm{t,substructure}}(\mathrm{r}_{\rm gc})=\sum \frac{\rm{d}^2\Psi_{\mathrm{sub-halo}}(\mathrm{r}_{\rm gc}-\mathrm{r^{\prime}})}{\rm{dq}^i \rm{dq}^j}$, where $\mathrm{r^{\prime}}$ is the location of each sub-halo. 

To determine how the sub-halo population will evolve with time and how frequently we need to calculate the tidal tensor, we first solve the orbit of each sub-halo using \texttt{galpy} and evolve each sub-halo forward for 10 Gyr. As a preliminary step, we first integrate sub-halo and cluster orbits at 10 Myr intervals to find the shortest interaction timescale that individual sub-halos have with three star cluster particles with circular orbits at 10 kpc, 50 kpc, and 100 kpc. We define the interaction timescale as the ratio of the relative distance to the relative velocity of each sub-halo with respect to each cluster. The orbit of each sub-halo and cluster is solved assuming it is a single body orbiting in a smooth isothermal potential with no substructure. We then reintegrate the sub-halo orbits and calculate the tidal tensor at the location of each test cluster at time intervals equal to $20\%$ of the shortest interaction timescale, ensuring that each close encounter that a cluster has with a sub-halo is fully resolved.

To determine how star cluster evolution depends on the fraction of substructure in a galaxy, we generate galaxy models with substructure fractions equal to 0$\%$, 1$\%$, 3$\%$ and 10$\%$. With \citet{penarrubia18} finding that the degree of fluctuation in the tidal field will depend on the actual structure of the perturbing sub-halos and their mass function, we also consider galaxy models where the assumed properties of the dark matter substructure are different. \citet{spitzer58} first showed that the timescale for shock dissolution is inversely dependent on the surface density of the perturber. The density profile of a perturbing sub-halo sets how much mass is interior to an interaction's impact parameter and, combined with the sub-halo's relative velocity, the shock strength. Hence in cases where the impact parameter is significantly less than a sub-halo's scale radius, dense sub-halos result in stronger tidal shocks than more extended sub-halos of equal mass and equal relative velocity. We therefore expect clusters will respond differently to sub-halos that follow a different mass-size relation than assumed above. In fact, \citet{penarrubia18} predicts that for a sufficiently steep mass-size relation that fluctuations in the tidal tensor (and therefore a cluster`s evolution) will be dominated by low-mass sub-halos since they will be both numerous and extremely dense compared to high-mass sub-halos. This argument also suggests that extending the sub-halo mass function to sub-solar masses may also lead to more fluctuation in the tidal tensor for select mass-size relations.

To test the predicted dependence of cluster evolution on substructure density, without having to simulate clusters in galaxy models with an array of mass-size relations, we simply consider the extreme case of sub-halos being point-masses. Evolving clusters in identical galaxy models, albeit with point-mass sub-halos, will result in more mass loss compared to the Hernquist sphere sub-halo galaxy models and set the upper limit to which denser sub-halos can affect cluster evolution (for the substructure mass function and fractions considered here). Point-mass sub-halos also serve as a proxy for interactions with compact sub-halos in general, as long as the entirety of a sub-halo's mass is within the interaction's impact parameter. As we will show in Section \ref{sec:hernquist}, Hernquist sphere sub-halos following our assumed mass-size relation already have a negligible effect on star clusters. Hence it is not necessary to test the effect of more extended sub-halos on cluster evolution.

To test how much low-mass sub-halos contribute to both the degree of fluctuation in the tidal tensor and cluster mass loss rates, we consider an additional set of galaxy models consisting of point-mass sub-halos where the lower sub-halo mass limit has been increased from $10^{5}$ to $10^{6} {\rm M}_\odot$. Hence clusters will undergo less tidal heating and fewer close encounters, which should minimize the effect that sub-halos have on cluster evolution. The new set of models were simulated assuming sub-halos are point-masses, as compact sub-halos have a much stronger effect on star clusters than Hernquist spheres and will better exemplify how a change in the sub-halo mass function affects star cluster evolution.

It should be noted that the masses, positions, and velocities of individual sub-halos are identical in galaxy models with a given $\rm{f}_{\mathrm{\rm sub}}$. When modelling galaxies with a narrower sub-halo mass range, we simply remove sub-halos with masses below the new minimum sub-halo mass and redistribute their mass into the smooth dark matter component of the galaxy. This approach ensures that model clusters in galaxies with a given $\rm{f}_{\mathrm{\rm sub}}$ are subjected to the exact same sub-halo interactions, which is especially important if a cluster undergoes a close interaction during its lifetime. It should also be noted that this approach will slightly reduce the actual value of $\rm{f}_{\rm sub}$ for the model galaxy.

\subsection{$N$-body Star Cluster Models}

To focus on how substructure affects star cluster evolution, the model star clusters are simply 50,000 stars of mass $0.6$ ${\rm M}_\odot$ that do not undergo stellar evolution. We model star clusters with circular orbits at 10 kpc, 50 kpc, and 100 kpc. The initial stellar distribution function is set equal to a Plummer sphere and the density profiles are scaled such that the ratio of the cluster's half-mass radius $\rm r_m$ to its tidal radius $\rm r_t$ is equal to either 0.145 or 0.245. Hence each cluster is tidally filling \citep{henon61} and will therefore be affected by the external tidal field. Clusters with $\rm r_m / r_t = 0.245$ are significantly less dense than clusters with $\rm r_m / r_t = 0.145$, which will result in them being more susceptible to tidal heating \citep{spitzer58}.

\begin{figure}
\centering
\includegraphics[width=\columnwidth]{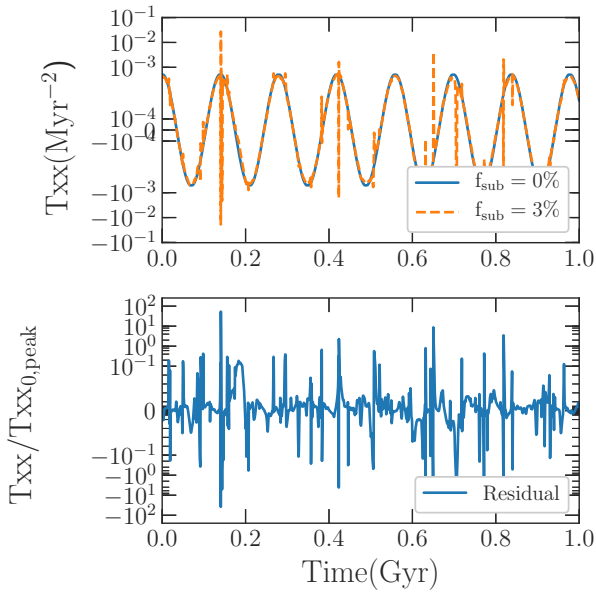}
\caption{Upper panel: $\mathbf{T}_{\mathbf{T}}^{xx}$ (the 0,0 component of the tidal tensor in a non-rotating reference frame) as a function of time for a cluster orbiting at 10 kpc in a galaxy model with point mass substructure fractions of 0 (black) and $3\%$ (red). Lower panel: The residual tidal tensor, which is the difference between the $3\%$ substructure case and the no-substructure case. Note that the y-axis in both panels has a symmetric log scale.}
  \label{fig:ttxx}
\end{figure}

Using \texttt{galpy}, the orbits of the clusters are solved beforehand (as discussed in Section \ref{sec:gal}) such that the tidal tensor can be provided to NBODY6tt as an input for the simulation with a time resolution equal to $20\%$ of the minimum interaction timescale experienced by the cluster. For illustrative purposes, we compare the first gigayear of the evolution in $\mathbf{T}_{\mathbf{T}}^{xx}$ for a cluster orbiting at 10 kpc in a galaxy with no substructure to the case of a galaxy with $3\%$ substructure in Figure \ref{fig:ttxx}. The normalized residual tidal tensor, where the smooth tensor is subtracted from the $3\%$ substructure tensor and then divided by the peak of the smooth tensor, is effectively the relative tidal tensor due to dark matter substructure only and is plotted in the lower panel of Figure \ref{fig:ttxx}.

Figure \ref{fig:ttxx} demonstrates that including substructure introduces two effects into a cluster's dynamical history; tidal heating due to long-duration, distant interactions and heating in the form shocks via short-duration, close interactions. Tidal heating due to distant interactions is reflected as minor fluctuations, on the order of $10\%$, in the tidal tensor about the mean background tidal field due to the halo simply not being a smooth distribution of dark matter. Both the frequency and intensity of the fluctuations will increase as a function of $\rm{f}_{\mathrm{sub}}$. Tidal shocks appear as sharp spikes in $\mathbf{T}_{\mathbf{T}}^{xx}$, which in Figure \ref{fig:ttxx} are up to twice as strong as the background tidal field, occurring when a sub-halo has a close encounter with the cluster. Both mechanisms serve to inject energy into the cluster, as fluctuations in the tidal tensor lead to an increase in the tidal heating parameter $\Delta \mathrm{E} \propto \mathrm{I_{tide}}=\sum (\int \mathrm{T_t}^{ij} \mathrm{dt})^2$ \citep{gnedin03} and are therefore expected to accelerate its dissolution  \citep{prieto08, kruijssen11}. However, as pointed out in \citet{spitzer58} and indicated by the definition of $\mathrm{I_{tide}}$, the duration of the shock is equally important as its strength. Hence a fast shock can have a minimal effect on a cluster if it happens over a short timescale. 


\subsection{Limits of the Tidal Approximation}\label{sec:approx}

As previously discussed, the tidal approximation is built around the assumption that a star's distance from the centre of the cluster is much less than the distance between the cluster and the centre of the external potential. While this is a reasonable assumption for clusters orbiting in a host galaxy, the accuracy of applying this approach to sub-halos passing nearby (and perhaps even through) a cluster is unclear. By definition the tidal approach will accurately predict the force acting on stars near the cluster's centre where the tidal tensor has been calculated, but will start to break down as a star's clustercentric distance increases or the distance between the cluster and the closest sub-halo decreases. 

\begin{figure}
\centering
\includegraphics[width=\columnwidth]{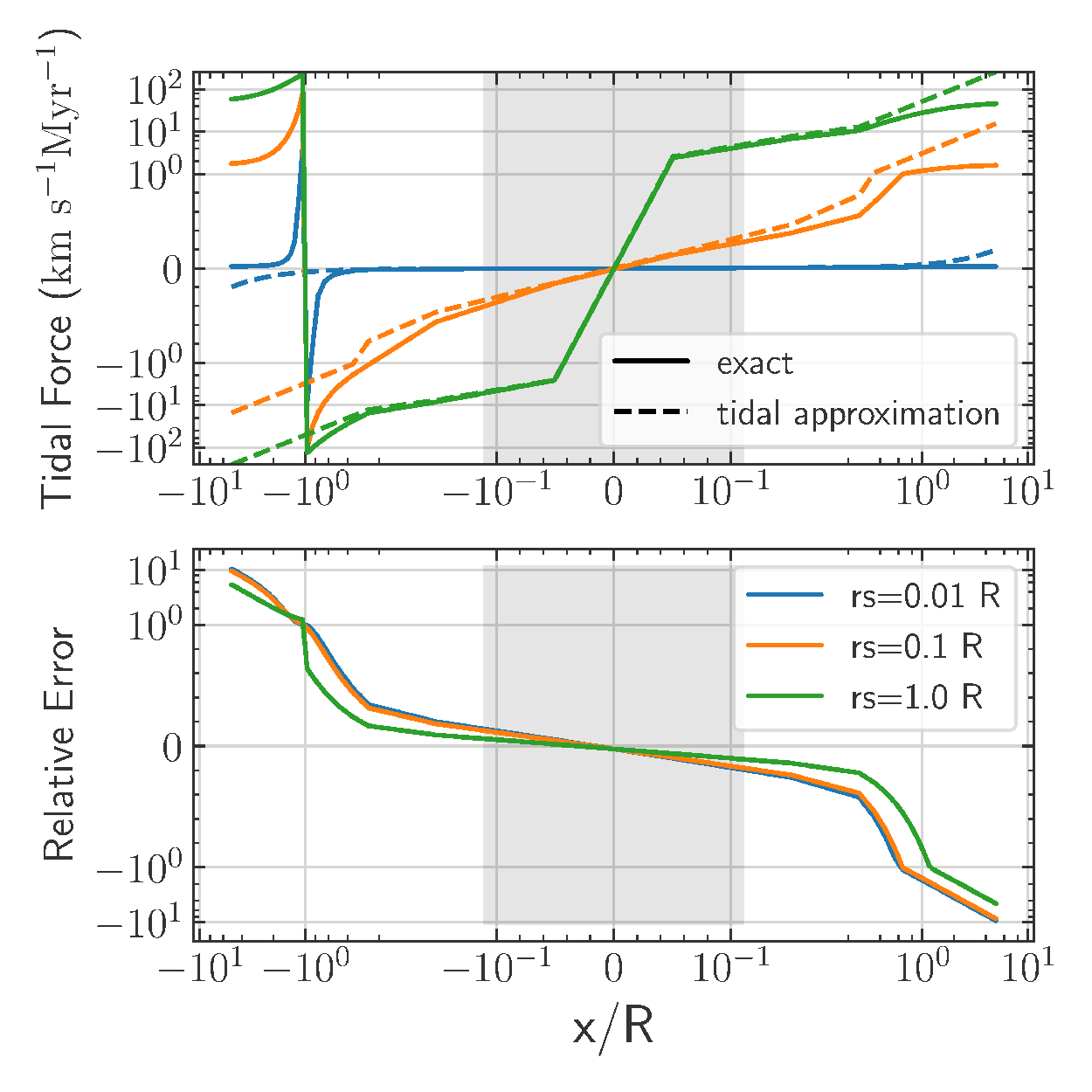}
\caption{Top panel: Tidal force due to a sub-halo located a distance R from a star cluster experienced by a star at a distance x from the cluster's centre. The tidal force has been calculated exactly (solid lines) and using the tidal approximation (dashed lines) for Hernquist sphere sub-halos with scale radii equal to 0.01R (blue), 0.1R (orange), and R (green) and assuming a 1/2 power-law mass size relation. The shaded grey region encompasses $99\%$ of the sub-halo - cluster interactions in our models (note the symmetric log x-axis). Bottom panel: Relative error between the exact tidal force and the tidal approximation.}
  \label{fig:shpot}
\end{figure}

To test the validity of the tidal approximation in the context of this study, we consider a star located along an axis connecting the centres of its host star cluster and a nearby sub-halo. We then determine the actual tidal force acting on the star due to a nearby Hernquist sphere sub-halo and the force calculated using the tidal approximation as a function of both the star's clustercentric distance and the sub-halo's distance from the cluster. For illustrative purposes, the upper panel of Figure \ref{fig:shpot} shows the tidal forces due to sub-halos with scale radii equal to $1\%$, $10\%$ and $100\%$ of the distance between the sub-halo and the cluster. The relative error is plotted in the lower panel Figure \ref{fig:shpot}, where the shaded region marks where $99\%$ of the interactions in our models occur.

Figure \ref{fig:shpot} illustrates that, for most of the sub-halo - cluster interactions in our models, the tidal approximation is capable of estimating the force acting on individual stars due to a sub-halo to within $30\%$. The tidal force acting on stars between the sub-halo and the cluster centre is slightly under-estimated while the force acting on stars on the opposite side of the cluster as the sub-halo is slightly over-estimated. The accuracy improves greatly for stars with smaller clustercentric distances. However for penetrating encounters, where the sub-halo is interior to the cluster's tidal radius (x/R $>=$ 1), the approximation begins to break down for stars near the sub-halo. In these cases, which are rare for the cluster and sub-halo properties considered here, the tidal approximation can differ from the actual tidal force by an order of magnitude if the sub-halo's scale radius is comparable to the size of the cluster. Furthermore, stars exterior to the sub-halo will have the tidal force over-estimated in the opposite direction, as the tidal approximation assumes the perturbing sub-halo is far from the host cluster. On the opposite side of the cluster as the sub-halo, the tidal force will also be over-estimated (but in the correct direction).

In general it appears that the tidal approach can accurately model the amount that outer stars are perturbed by nearby sub-halo passes, but may overestimate the effect of sub-halos which pass through the cluster. Given the model galaxies considered in this study, no cluster is expected to have a sub-halo pass through its half-mass radius within 10 Gyr. Extending the interaction cross-section to consider sub-halos passing through each cluster's tidal radius, we expect on the order of 1 direct interaction in galaxies with substructure fractions of $1\%$ and $3\%$ and on the order of 10 direct interactions in galaxies with substructure fractions of $10 \%$. These encounter rates decrease even further if we require the sub-halo to be similar in size to the cluster. Therefore we conclude that in general using the tidal approach will slightly underestimate the effect that dark matter substructure has on star cluster evolution, as underestimating the force acting on outer regions stars due to nearby encounters will lead to an underestimation of each cluster's mass loss rate. However in the rare occurrence of a direct encounter with a sub-halo of equal or lesser size, the tidal approximation will overestimate the amount of mass lost by the clusters. 

\section{Results} \label{s_results}

To study the effects that dark matter sub-halo interactions have on star clusters, we first explore how both the structural and kinematic properties of model clusters in each of the galaxy models with Hernquist sphere sub-halos evolve with time relative to model clusters in galaxies that contain no substructure. A Hernquist sphere with a power-law mass-size relation is one of the more commonly assumed forms for dark matters sub-halos, used in both cosmological simulations and studies of how sub-halos interact with stellar streams. To probe how denser sub-halos can affect star cluster evolution, we also consider the extreme case where sub-halos are instead point-masses. Point-mass sub-halos will have the strongest effect possible on cluster evolution, which allows us to place upper limits on how much mass loss can be attributed to sub-halo interactions. It is important to note that in our analysis we only consider stars within the cluster's tidal radius $\rm{r_t}$ to be cluster members, where $\rm{r_t}$ is calculated using the formalism of \citet{bertin08} within \texttt{galpy}.  We iterate twice over the $\rm{r_t}$ calculation, meaning $\rm{r_t}$ is initially calculated using the mass of all stars in the simulation, with the final value of $\rm{r_t}$ then calculated using only the stars with clustercentric distances within this first calculation of $\rm{r_t}$. Since stars beyond $20 \times$ the cluster's half-mass radius are removed from the simulation completely, one iteration is typically enough for the calculation to converge except for when the cluster is near complete dissolution.


\subsection{Hernquist Sphere Sub-halos}\label{sec:hernquist}

\begin{figure*}
\centering
\includegraphics[width=\textwidth]{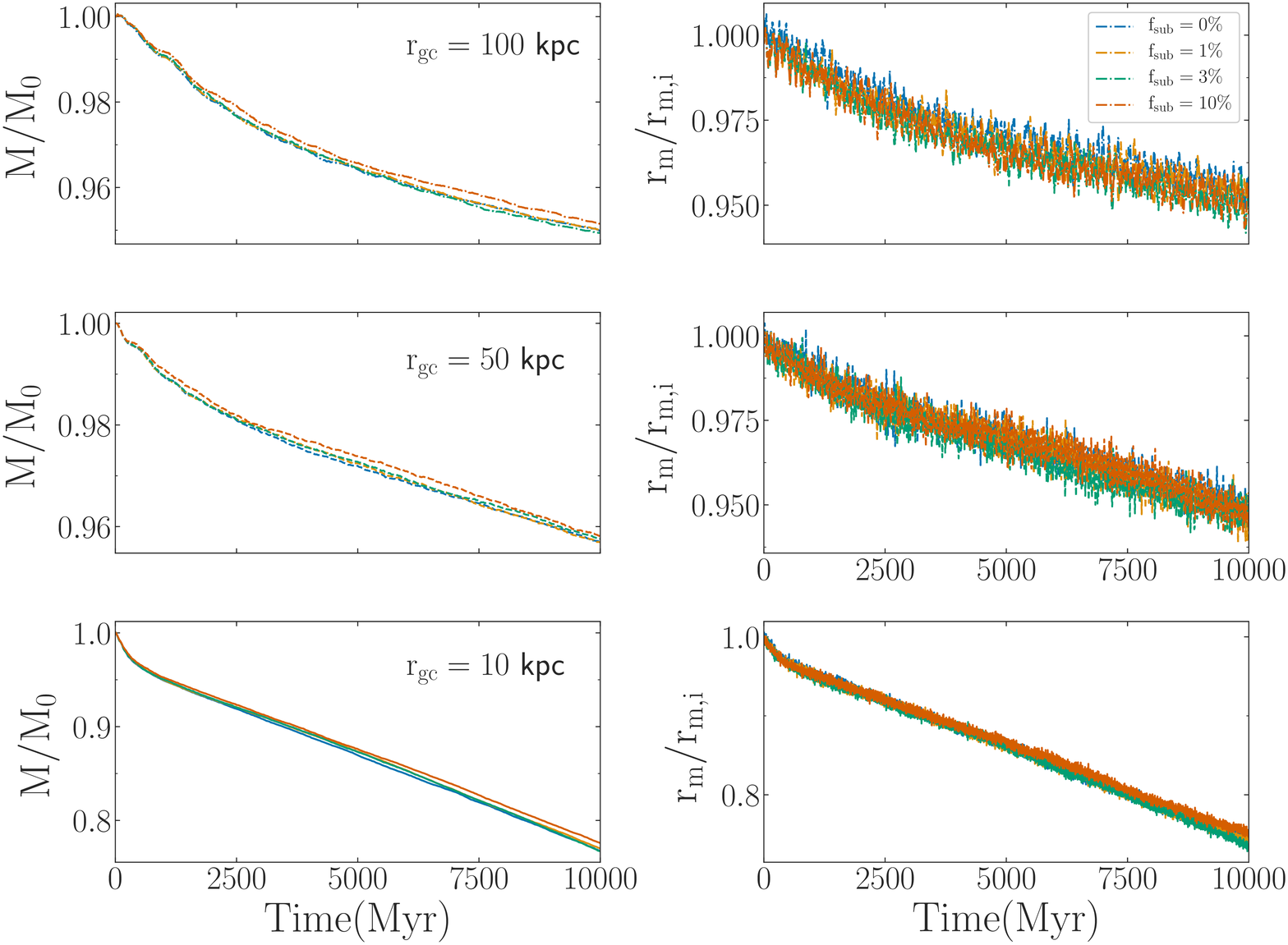}
\caption{Left Panel: Mass fraction as a function of time for clusters with $\rm r_m / r_t = 0.145$  orbiting at a galactocentric distance of 10 kpc (bottom), 50 kpc (middle), and 100 kpc (top) in galaxy models with Hernquist sphere substructure fractions of $0\%$ (blue), $1\%$ (orange), $3\%$ (green) and $10\%$ (red). Right Panel: Same as left panel but for cluster half mass radii.}
  \label{fig:hplot}
\end{figure*}

Under the assumption that dark matter sub-halos take the form of Hernquist spheres with a mass-size relation of the form $\rm{r_s}=1.05$ kpc $(\rm{M}/10^8 {\rm M}_\odot)^{\frac{1}{2}}$ and have masses between $10^5$ and $10^{11} {\rm M}_\odot$, we find that dark matter substructure has very little effect on the long term evolution of star clusters. In the left panel of Figure \ref{fig:hplot}, we illustrate the mass evolution of clusters orbiting at 10, 50, and 100 kpc with $\rm r_m / r_t = 0.145$ in model galaxies with substructure fractions between 0$\%$ and $10\%$. In all cases, it appears that dark matter substructure has a negligible effect on cluster evolution over the sub-halo mass range considered here. Even when considering lower density clusters with $\rm r_m / r_t = 0.245$, which are more susceptible to mass loss via tidal shocks than denser clusters \citep{spitzer58}, there is no difference between the mass evolution of clusters in model galaxies with and without substructure. The same can be said about the evolution of cluster half-mass radii, as illustrated in the right panel of Figure \ref{fig:hplot}.

Despite the presence of substructure causing clear fluctuations in the tidal tensor experienced by each model cluster, Figure \ref{fig:hplot} shows that clusters are not strongly affected by the presence of Hernquist sub-halos with the mass function and size-mass relation initially considered here. Hence the tidal tensor fluctuations due to distant sub-halo interactions are too weak and shocks due to close encounters are both too short and too weak to inject enough energy into the clusters to unbind stars. A closer look at the evolution of each cluster's velocity dispersion and anisotropy profiles shows, as expected, that no signature of dark matter substructure interactions exists in the kinematic properties of the model clusters either. After 10 Gyr evolution, the velocity dispersion and anisotropy profiles of each model cluster are identical.

\subsection{Point-Mass Sub-halos}\label{sec:a10}

Tidal fluctuations due to dark matter substructure are expected to be strongly dependent on the sub-halo mass-size relation, as denser sub-halos will have a stronger effect on cluster evolution than extended sub-halos \citep{spitzer58, penarrubia18}. For a given sub-halo mass, a smaller scale radius increases a sub-halo's contribution to $\mathbf{T}^{ij}_{\rm{t,substructure}}(\rm{r})$, which will result in a higher degree of tidal fluctuations than the Hernquist spheres considered in the previous section. To test the effects of denser sub-halos on cluster evolution, we make the extreme assumption that all sub-halos are point-masses instead of Hernquist spheres. While the effects of point mass sub-halos and compact Hernquist spheres on cluster evolution will be similar, they will be maximized in the point mass case.

\begin{figure*} 
\centering
\includegraphics[width=\textwidth]{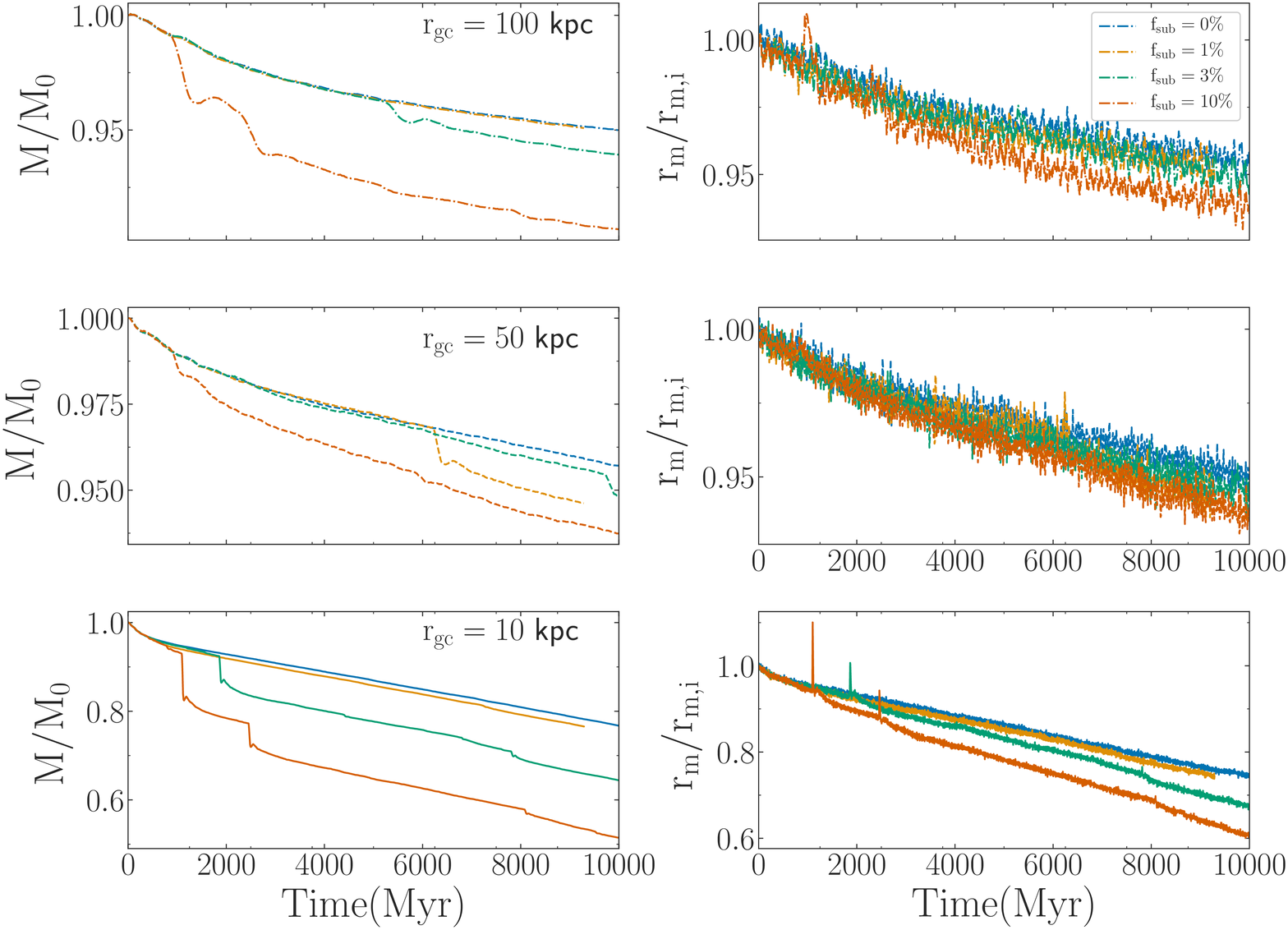}
\caption{Same as Figure \ref{fig:hplot}, but for point mass sub-halos.}
  \label{fig:kplot}
\end{figure*}

The left panel of Figure \ref{fig:kplot} illustrates the evolution of cluster mass as a function of time for clusters orbiting at 10, 50, and 100 kpc in galaxies with point mass substructure fractions of 0$\%$, 1$\%$, 3$\%$, and 10$\%$. For a given galactocentric distance, cluster mass loss rate increases as a function $\rm{f}_{\mathrm{\rm sub}}$. At early times the differences in mass loss rates are the result of distant tidal heating effects, which yields a small difference between model clusters. However as time goes on, sharp drops in mass occur due to tidal shocks induced by close encounters. Since the sub-halo encounter rate increases as a function of $\rm{f}_{\mathrm{\rm sub}}$, clusters in galaxies with larger substructure fractions end up losing mass at higher rates. For example, at 10 kpc the $\rm{f}_{\mathrm{\rm sub}}=1\%$ model cluster undergoes no significant close interactions within a Hubble time and is only subject to heating via distant sub-halo interactions, which results in the cluster being less than $1\%$ less massive than if it resided in a smooth dark matter halo. On the other hand, the $\rm{f}_{\mathrm{\rm sub}}=10\%$ model cluster at 10 kpc undergoes at least 3 strong encounters and several weak encounters which in addition to tidal heating leaves it with a final mass that is $\sim25\%$ less massive than the $\rm{f}_{\mathrm{\rm sub}}=0$ case. However, we caution that the tidal approximation likely overestimates the amount of mass loss during very close counters (see Section \ref{sec:approx}).

At 50 and 100 kpc, the influence of point mass dark matter sub-halos is clearly less important due to the local dark matter density being lower. In fact, it appears tidal heating from distant encounters is near negligible as the mass evolution of clusters in galaxies with $\rm{f}_{\mathrm{\rm sub}}$ values of $0\%$, $1\%$,  and $3\%$ are very similar until a cluster undergoes a close encounter with a point mass sub-halo. Statistically, such events become rarer as $\rm{f}_{\mathrm{\rm sub}}$ decreases or galactocentric distance increases. However it also appears that tidal heating in the $\rm{f}_{\mathrm{\rm sub}}=10\%$ case plays a role at both 50 and 100 kpc, as the model clusters are $\sim 2\%$ less massive than the corresponding $\rm{f}_{\rm sub}=0\%$ model after 10 Gyr. 

It is important to note that the dependence found here between how strongly clusters are affected by sub-halos and galactocentric distance is a bi-product of point-mass sub-halos not having a constant surface density as a function of mass. In the tidal approximation, the mass-loss timescale as the result of shocks from sub-halos can be expressed as: $t_{\rm sh} \propto \rho_{\rm h}\sigma_{\rm sub}/(\Sigma_{\rm sub}f_{\rm sub}\rho_{\rm gc})$ \citep{spitzer58}, where $\rho_{\rm h}$ is the average cluster density within $r_{\rm h}$, $\sigma_{\rm sub}$ is the velocity dispersion of the sub-halos, $\Sigma_{\rm sub}$ is the surface density of individual sub-halos and $\rho_{\rm gc}(r_{\rm gc})$ is the total DM density at $r_{\rm gc}$. When $\Sigma_{\rm sub}$ is constant (which is the case for our Hernquist sphere sub-halo models), Roche-filling clusters in a singular isothermal halo will have $\rho_{\rm h}\sigma_{\rm sub}/\rho_{\rm gc}=$constant. Hence $t_{\rm sh}$ is independent of $r_{\rm gc}$. The mass-loss timescale as the result of evaporation in the smooth tidal field depends on $r_{\rm gc}$ as $t_{\rm ev}\propto r_{\rm gc}$  for the singular isothermal sphere \citep{baumgardt03}. The ratio of $t_{\rm sh}/t_{\rm ev}\propto r_{\rm gc}^{-1}$ meaning that the effect of sub-halo interactions on the mass evolution of clusters is expected to be more important at large $r_{\rm gc}$ when sub-halos have a constant surface density. It is worth noting that the Hernquist sphere galaxy models considered here predict shock dissolution timescales of over $20 \times$ the age of the Universe for a standard tidally filling cluster (consistent with Figure \ref{fig:hplot}). However given that $t_{\rm sh} \propto \Sigma_{\rm sub}^{-1} f_{\rm sub}^{-1}$, higher substructure fractions and denser sub-halo's can yield dissolution times due to sub-halo interactions that are less than a Hubble time. 


To examine how cluster structure is affected by point mass substructure interactions, we also consider the evolution of each model cluster's half-mass radius in the right panel of Figure \ref{fig:kplot}. Since the model clusters are tidally filling, a tidal shock leads to stars escaping the cluster. Hence as the model clusters lose mass they decrease in size as well, with the exception of short lived episodes of expansion right after a close encounter, evidence of which is removed after the energized stars escape the cluster. For the 10 kpc model the difference in cluster size due to $1\%$ substructure is minimal. However when $\rm{f}_{\mathrm{\rm sub}}=10\%$ the model cluster is nearly $20\%$ smaller than the model cluster orbiting in a galaxy with $\rm{f}_{\mathrm{\rm sub}}=0\%$. At 50 kpc, where the local dark matter density is significantly lower, the difference between the $\rm{f}_{\mathrm{\rm sub}}=0\%$ and $\rm{f}_{\mathrm{\rm sub}}=10\%$ models reduces to just $3\%$. Finally at 100 kpc, differences in cluster size are at most $3\%$, which would be difficult to observe and to attribute the difference to sub-halos as opposed to the cluster's initial conditions. 

A closer look at each model cluster's density profile revealed no clear dependence on tidal shocks. In all cases cluster density decreases smoothly out to the extent of the cluster. While the decrease in cluster mass and size associated with an increase in $\rm{f}_{\mathrm{\rm sub}}$ is clearly visible, again there is no clear sign that the resulting density profile is due to tidal shocks and heating or simply because the cluster was smaller and less massive at birth.

\begin{figure*}
\centering
\includegraphics[width=\textwidth]{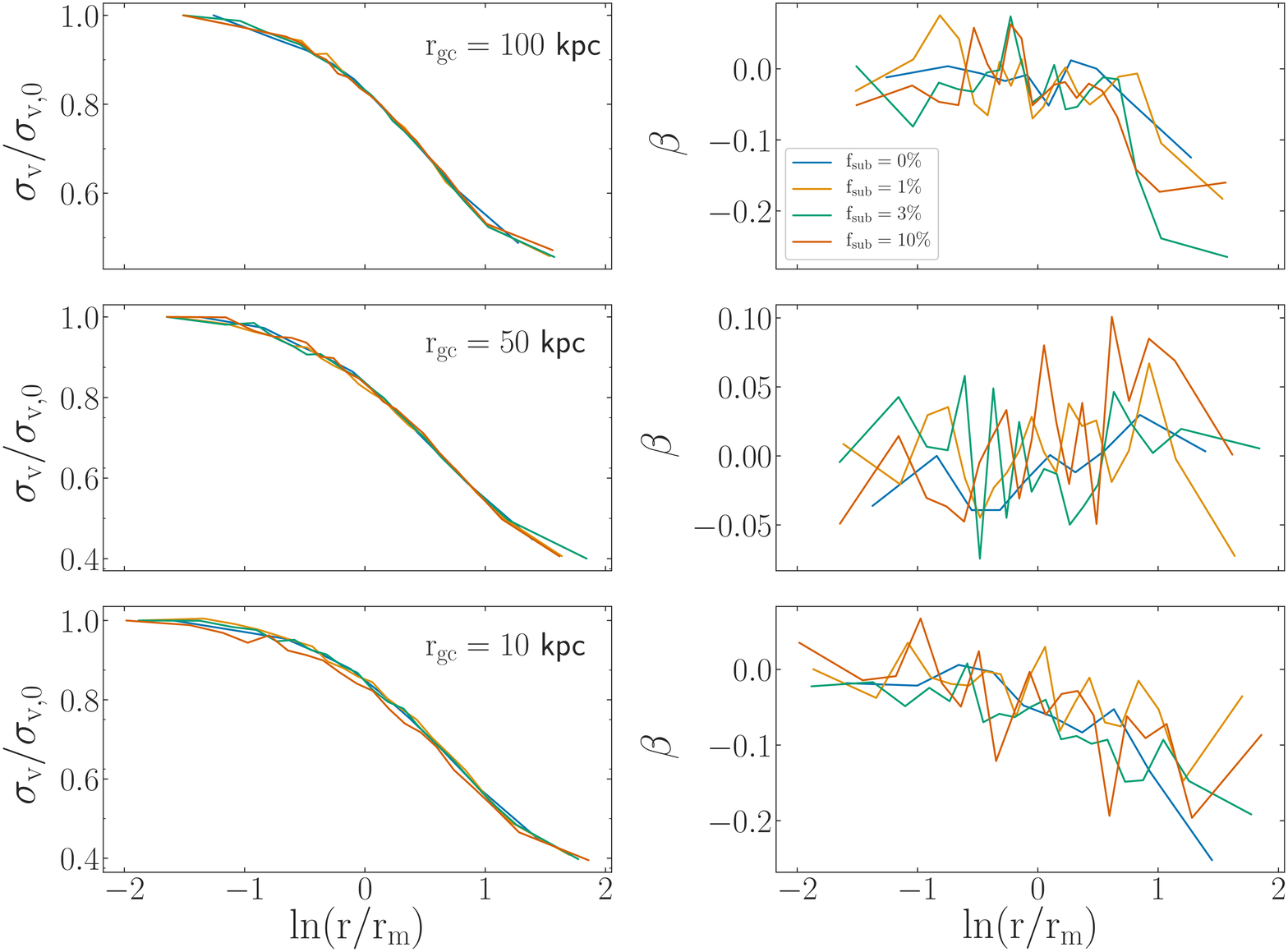}
\caption{Left panel: Velocity dispersion $\sigma_{\rm v}$ (normalized by the inner most value $\sigma_{\rm v,0}$)  as a function of the natural logarithm of clustercentric distance (normalized by the half-mass radius) for clusters orbiting at 10 kpc (bottom), 50 kpc (middle), and 100 kpc (top) in galaxy models with different point mass substructure fractions after 10 Gyr. Right panel: Same as left panel but for the orbital anisotropy parameter $\beta$.}
  \label{fig:beta}
\end{figure*}

To explore how point mass dark matter sub-halo interactions affect star cluster kinematics, we also considered the evolution of both the three-dimensional velocity dispersion profile and the orbital anisotropy profile of the cluster. Similar to the model cluster's density profiles, the velocity dispersion profiles of clusters in model galaxies with substructure do not differ greatly from model galaxies without, with two exceptions. The first exception occurs immediately after a sub-halo encounter, when a tidal shock causes the velocity dispersion in the outer region of the cluster to increase significantly for approximately one crossing time before returning to follow a smooth decrease with clustercentric distance as expected. Repeated tidal shocks lead to the second exception, which is illustrated in the left panel of Figure \ref{fig:beta} where we show the velocity dispersion profile of each model cluster after 10 Gyr of evolution. Figure \ref{fig:beta} shows a weak but clear trend in the behaviour of the velocity dispersion of stars in the outermost region of a cluster, with the outer velocity dispersion increasing as a function of the dark matter substructure fraction for a given orbital distance. This increase is a direct result of clusters being subject to tidal heating due to substructure interactions. The difference between models is only on the order of a few percent, and would be difficult to measure observationally.

A complementary signature also exists in the anisotropy profile of each model cluster. Similar to the velocity dispersion profile, the anisotropy parameter $\beta$ sharply increases towards 1 (radial anisotropy) after a tidal shock. Also like the evolution of the velocity dispersion profile, this signal disappears within a crossing time as stars energized to radial orbits quickly escape the cluster. Over 10 Gyr of evolution, a trend does begin to emerge with $\rm{f}_{\mathrm{\rm sub}}$ after clusters in weak tidal fields have been subjected to tidal heating for a long period of time. As shown in the right panels of Figure \ref{fig:beta}, after 10 Gyr $\beta$ steadily decreases with clustercentric distance as high-eccentricity stars are more easily removed from the cluster with time. In the lower panel of Figure \ref{fig:beta}, we see that $\beta$ profiles appear to be independent of $\rm{f}_{\mathrm{\rm sub}}$ for clusters orbiting at 10 kpc and 50 kpc. Hence outer region stars that receive an energy boost due to tidal heating and shocks become unbound regardless of their orbital eccentricity due to the cluster being subject to a strong background tidal field. At 100 kpc on the other hand, where the external tidal field is weaker, it is clear that $\beta$ decreases in the outer regions of clusters that are subject to sub-halo interactions. Thus stars that are primarily on radial orbits are energetic enough to escape the cluster when they are heated, leaving only stars with tangential orbits in the outskirts of the cluster. The fact that a cluster's velocity dispersion and anisotropy profiles are affected by sub-halos offers a potential observational signature of dark matter substructure interactions within globular clusters, however again the signature is small and difficult to observe. Furthermore, a large suite of simulations is necessary to determine the exact range of sub-halo mass functions and mass-size relations that can sufficiently alter cluster kinematics. Detailed dynamical modelling of stars clusters \citep[e.g.][] {zocchi16,peuten17,henaultbrunet19} may offer the best chance of observing the kinematic signature of clusters interacting with sub-halos, as the behaviour of $\sigma_{\rm v}$ and $\beta$ at large cluster centric distances will be difficult to reproduce.

Ultimately, studying the effects that point-mass sub-halos have on star clusters serves as in indication of how sensitive star clusters are to the sub-halo mass-size relation. The models demonstrate that cluster evolution will be sensitive to the amplitude and slope of the mass-size relation, especially in the $\rm{f}_{\mathrm{\rm sub}}=10\%$ models where interactions with substructure are more frequent. A steeper or scaled sub-halo mass-size relation that results in Hernquist sphere sub-halos being more compact than those discussed in Section \ref{sec:hernquist} would lead to an increase in cluster disruption. Hence star clusters may be able to constrain the sub-halo mass-size relation in the Milky Way in a similar fashion to how \citet{moore93} used the existence of several Galactic GCs to rule out dark matter particles being black holes with masses greater than $1000 {\rm M}_\odot$. For lower substructure fractions, substructure already has a minimal effect on cluster evolution such that differences due to the size of individual sub-halos will be negligible.

\section{Discussion}\label{s_discussion}

We have modelled the evolution of tidally filling star clusters in a logarithmic galaxy potential with a range of dark matter substructure fractions and properties. The effects of Hernquist sphere sub-halos, the density profile commonly attributed to sub-halos in the CDM framework, have proven to be negligible with respect to a cluster's mass and size evolution for the sub-halo mass range and mass-size relation considered here. Only point-mass sub-halo interactions, or potentially Hernquist sphere sub-halos that are extremely compact (due to either a change in the scaling factor or the power-law dependence of the mass-size relation) can affect cluster evolution. 


To understand why cosmologically motivated Hernquist sphere sub-halos have no effect on cluster evolution, despite clearly affecting the tidal tensor, we explore how cluster mass loss rates, $\rm{f}_{\rm sub}$, and fluctuations in the tidal tensor are all related for both Hernquist sphere and point-mass sub-halos. As a proxy for the degree of fluctuation in each tidal tensor we calculate the square root of the sum of the squares of the tensor's eigenvalues ($\Lambda=\sqrt{\sum \lambda_i^2}$) at each time-step and calculate both the standard deviation ($\sigma_{\rm TT}$) and the mean absolute deviation ($\sigma_{\rm MAD,TT}$) of $\Lambda$. It is informative to calculate both $\sigma_{\rm TT}$ and $\sigma_{\rm MAD,TT}$, as $\sigma_{\rm TT}$ is more sensitive to the tails of a distribution than $\sigma_{\rm MAD,TT}$. Hence $\sigma_{\rm TT}$ provides an indication of the degree in which a cluster is subject to tidal shocks via close encounters, while $\sigma_{\rm MAD,TT}$ traces minor fluctuations in the tidal tensor due to more distant sub-halos.

\subsection{Relating Cluster Mass Loss to the Tidal Tensor}

\begin{figure}
\centering
\includegraphics[width=\columnwidth]{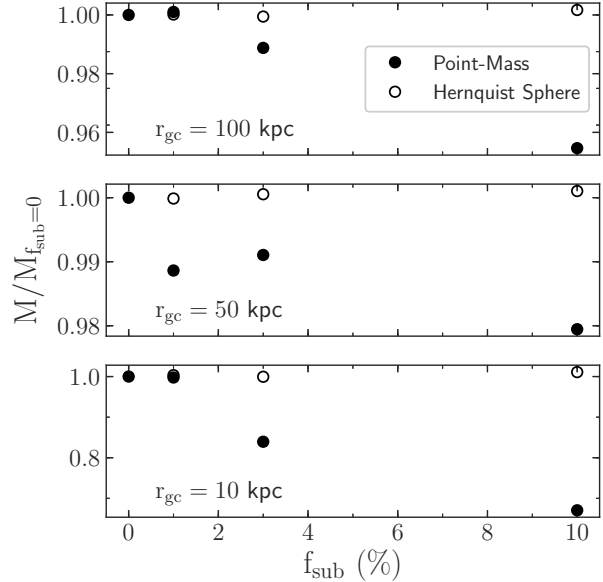}
\caption{Fraction of mass lost compared to the no-substructure case ($M/M_{\rm{f}_{\rm sub}=0}$)  as a function of $\rm{f}_{\mathrm{\rm sub}}$ for clusters orbiting in galaxy models with point-mass sub-halos (filled circles) and Hernquist sphere sub-halos (open circles) at a galactocentric distance of 10 kpc (bottom), 50 kpc (middle) and 100 kpc (top).}
  \label{fig:mfplot}
\end{figure}

We first explore how the fraction of mass lost by each model cluster depends on $\rm{f}_{\mathrm{\rm sub}}$ in Figure \ref{fig:mfplot}, where the ratio between the mass of each model cluster after 10 Gyr of evolution relative to the $\rm{f}_{\mathrm{\rm sub}}=0$ model is plotted for clusters orbiting at 10 kpc, 50 kpc, and 100 kpc in the galaxy models with point-mass and Hernquist sphere sub-halos. While M/M$_0$ is expectedly constant at 1 for Hernquist sphere sub-halos due to their negligible effect on cluster evolution, we find that M/M$_0$ is inversely proportional to $\rm{f}_{\mathrm{\rm sub}}$ for point-mass sub-halos as the fraction of mass lost increases with $\rm{f}_{\mathrm{\rm sub}}$. The approximate linearity of the relationship is consistent with the derivation of \citet{spitzer58} (and more recently by \citealt{gieles06}), who finds that cluster mass loss rate depends linearly on the local density of perturbing giant molecular clouds based on the impulse approximation.

\begin{figure*}
\centering
\includegraphics[width=\textwidth]{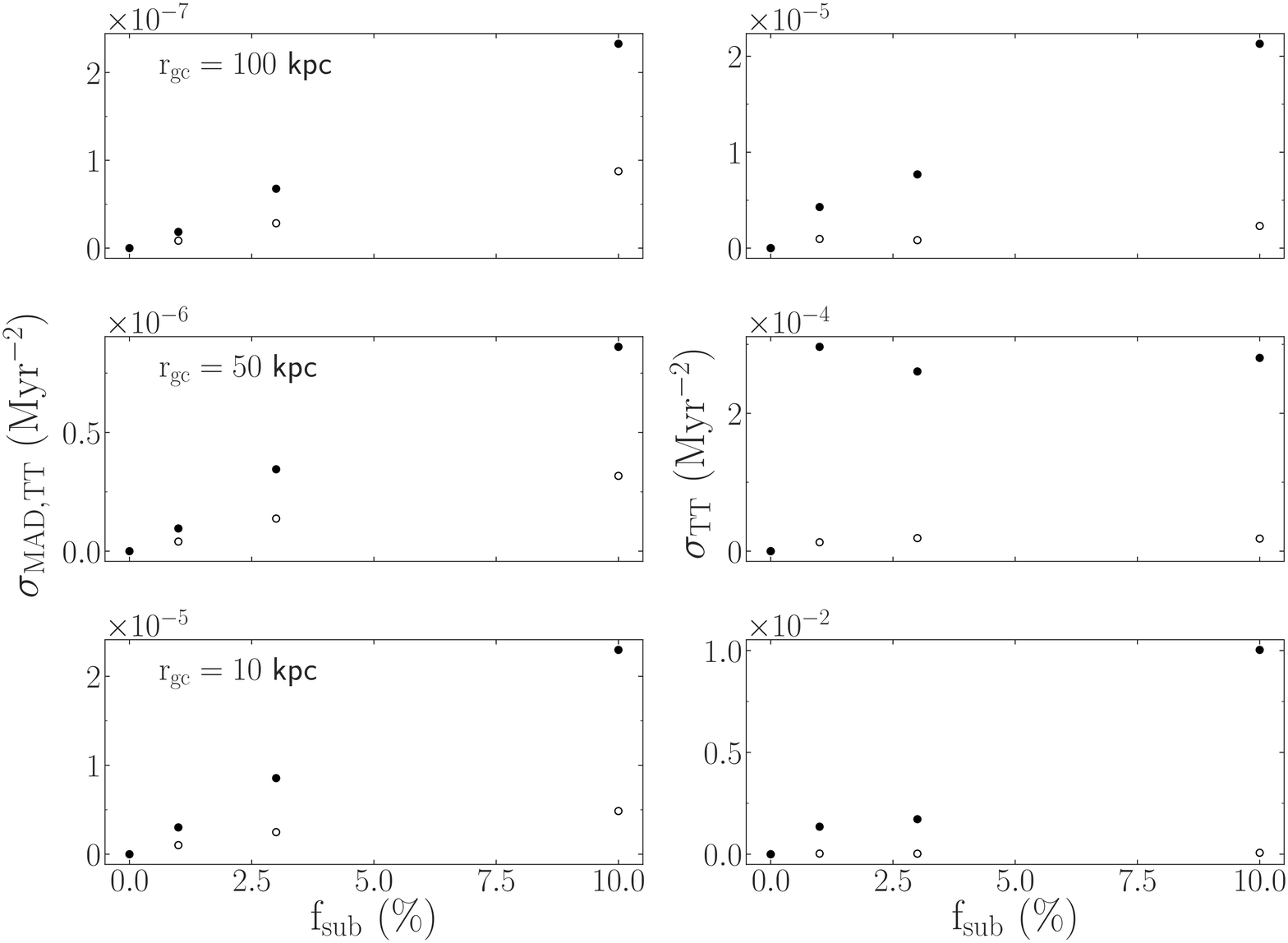}
\caption{Left Panel: Mean absolute deviation of the sum of the squares of the tidal tensor's eigenvalues due to dark matter substructure as a function of $\rm{f}_{\mathrm{\rm sub}}$ for clusters orbiting in galaxy models with point-mass sub-halos (filled circles) and Hernquist sphere sub-halos (open circles) at a galactocentric distance of 10 kpc (bottom), 50 kpc (middle) and 100 kpc (top). Right Panel: Same as the left panel but for the standard deviation of the sum of the squares of the tidal tensor's eigenvalues due to dark matter substructure.}
  \label{fig:sigttf}
\end{figure*}

For a given $\rm{r}_{\rm gc}$, the relationship between M/M$_0$ and $\rm{f}_{\mathrm{\rm sub}}$ is expected to be directly related to fluctuations in the tidal tensor as it is the only thing that varies between simulations. As previously discussed, we quantify fluctuations in the tidal tensor with $\sigma_{\rm TT}$ and $\sigma_{\rm MAD,TT}$, both of which are plotted for galaxy models with point-mass and Hernquist sphere sub-halos in Figure \ref{fig:sigttf} (with clusters separated by $r_{\rm gc}$). As expected, both $\sigma_{\rm MAD,TT}$ and $\sigma_{\rm TT}$ increase with $\rm{f}_{\mathrm{\rm sub}}$ at a given $r_{\rm gc}$, with both values being larger in general when clusters orbit in denser environments (low $r_{\rm gc}$). When comparing the $\sigma_{\rm MAD,TT}$ and $\sigma_{\rm TT}$ of Hernquist sphere sub-halos to point mass sub-halos, both values are between 5 and 10 times higher for point mass sub-halos, which corresponds to the significant increase in cluster mass loss rates observed when modelling dark mark sub-halos as point masses.

\begin{figure}
\centering
\includegraphics[width=\columnwidth]{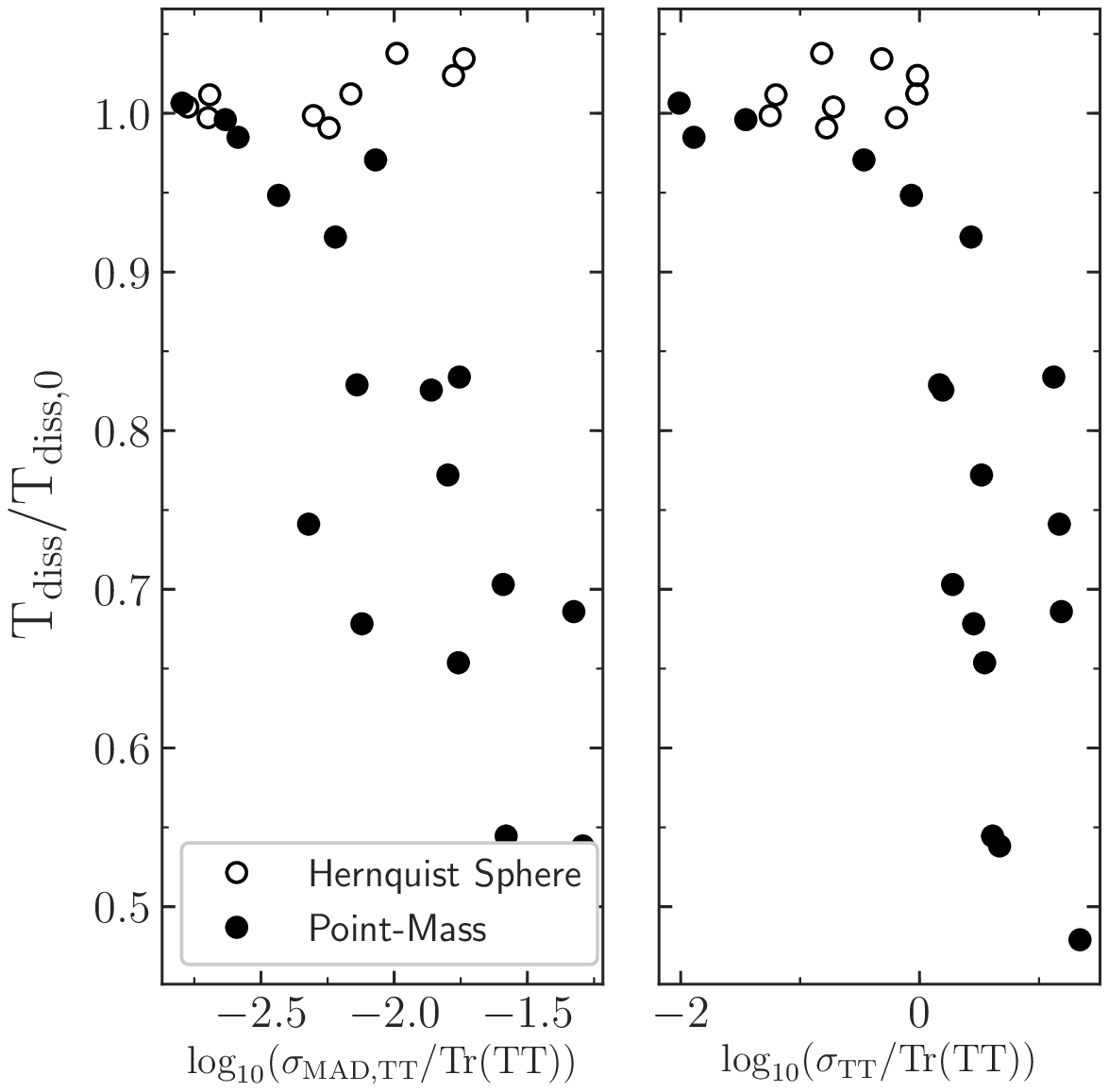}
\caption{Ratio of cluster dissolution time to the dissolution of a cluster at the same $r_{\rm gc}$ but with $\rm{f}_{\mathrm{\rm sub}}=0$ as a function of $\log_{10} (\sigma_{\rm MAD,TT}/\rm{Tr}(\rm{TT}))$ (left panel) and $\log_{10} (\sigma_{\rm TT}/\rm{Tr}(\rm{TT}))$ (right panel). Filled circles and open circles represent point mass and Hernquist sphere sub-halos respectively.}
  \label{fig:tdissn}
\end{figure}

Figures \ref{fig:mfplot} and \ref{fig:sigttf} suggest that the degree of fluctuation in the tidal tensor the cluster experiences, a cluster's galactocentric distance, and the local value of $\rm{f}_{\rm sub}$ all play a role in determining a cluster's mass loss rate. Therefore we calculate the ratios $\sigma_{\rm MAD,TT}/\rm{Tr}(\rm{TT})$ and $\sigma_{\rm TT}/\rm{Tr}(\rm{TT})$ for each tidal tensor history (where Tr(\rm{TT}) is the trace of the tidal tensor associated with each model cluster) to compare with each cluster's mass loss rate. The trace of the tidal tensor is proportional to the background dark matter density, which in turn probes the strength of the background tidal field. In the left panel of Figure \ref{fig:tdissn} we plot the ratio of each model cluster's dissolution time to the dissolution time of a model cluster orbiting at the same $r_{\rm gc}$ but with $\rm{f}_{\rm sub}=0$ as a function of $\log_{10} (\sigma_{\rm MAD,TT}/\rm{Tr}(\rm{TT}))$. Dissolution times are estimated by taking the mean mass loss rate of each model cluster over 10 Gyr. In the right panel of Figure \ref{fig:tdissn} we instead calculate the ratio $\log_{10}  (\sigma_{\rm TT}/\rm{Tr}(\rm{TT}))$. 

Figure \ref{fig:tdissn} illustrates that for Hernquist sphere sub-halos, cluster dissolution times will remain unchanged for a wide range in $\log_{10} (\sigma_{\rm MAD,TT}/\rm{Tr}(\rm{TT}))$ and $\log_{10} (\sigma_{\rm TT}/\rm{Tr}(\rm{TT}))$. Point mass sub-halos, on the other hand, show a weak dependence on both $\log_{10} (\sigma_{\rm MAD,TT}/\rm{Tr}(\rm{TT}))$ and $\log_{10} (\sigma_{\rm TT}/\rm{Tr}(\rm{TT}))$. A separation between point-mass and Hernquist sphere sub-halos also appears to exist at $\log_{10} (\sigma_{\rm TT}/\rm{Tr}(\rm{TT})) = 0$, with the Hernquist sphere sub-halo's considered here not being able to generate fluctuations high enough to reach such large values of $\sigma_{\rm TT}$. Since $\sigma_{\rm TT}$ is more sensitive to close encounters than $\sigma_{\rm MAD,TT}$, it is likely that tidal shocks due to close encounters are the key driver behind the accelerated mass loss of clusters subject to dark matter substructure interactions and the scale radius of Hernquist sphere sub-halos minimizes the strength of such shocks. In fact, differences between the point-mass sub-halos and Hernquist sphere sub-halos in Figures \ref{fig:mfplot} and \ref{fig:tdissn} suggest that tidal shocks must surpass a threshold value of $\sigma_{\rm TT}$, which likely depends on both the cluster's density and the local tidal field, before resulting in cluster mass loss.

In order to test the importance of short-duration close encounters relative to long-duration distant sub-halo encounters, we re-simulate model clusters in galaxies with point-mass sub-halos but in the idealized cases of suppressing encounters that induce fluctuations in the tidal tensor that are either greater than or less than $20\%$ of the smooth tidal tensor's maximum amplitude. These two scenarios effectively separate the effects of short-duration close encounters and long-duration distant sub-halo encounters. From these additional simulations we find that close encounters are responsible for almost the entire amount of additional mass loss due to interactions with sub-halos, as tensors where encounters that cause fluctuations greater than $20\%$ of the smooth tidal tensor are suppressed yield clusters that are nearly identical to clusters orbiting in model galaxies with no substructure. Therefore it is necessary for Hernquist sphere sub-halos to be dense enough such that close encounters cause $\log_{10} (\sigma_{\rm TT}/\rm{Tr}(\rm{TT})) > 0$ in order to affect star cluster evolution.

In general, for $\log_{10} (\sigma_{\rm TT}/\rm{Tr}(\rm{TT})) > 0$  the point-mass sub-halo models weakly suggest that  $T_{diss}/T_{diss,0}$  $\propto$ $\log_{10} ((\sigma_{\rm TT}/\rm{Tr}(\rm{TT}))^{-1/8})$, with a slightly steeper (-1/3) trend observed with respect to $\log_{10} (\sigma_{\rm MAD,TT}/\rm{Tr}(\rm{TT}))$. Significant scatter is expected about the relation for two reasons. First, a model cluster that has recently experienced a close encounter and a burst of mass lost will have a higher than expected mass loss rate and shorter dissolution time compared to the mean values over the course of its lifetime. Similarly a model cluster that has yet to undergo a close encounter, or hasn't in awhile, will have an underestimated dissolution time. Second, close encounters between star clusters and sub-halos that create a large fluctuation in the tidal tensor can sometimes have a minimal effect on stars in the cluster if the encounter timescale is much shorter than the cluster's crossing time. Hence some fluctuations in the tidal tensor do not result in mass loss. 

This second point is reflected in the differences between clusters orbiting in galaxy models with Hernquist sphere sub-halos and point-mass sub-halos. In some of the Hernquist sphere cases, which do not accelerate cluster dissolution times, both  $\log_{10} (\sigma_{\rm MAD,TT}/\rm{Tr}(\rm{TT}))$ and $\log_{10} (\sigma_{\rm TT}/\rm{Tr}(\rm{TT}))$ are so large that they are comparable to the point mass sub-halo models which have dissolution times that are half of the no-substructure case. However most of these encounters occur with sub-halos and GCs having high relative speeds, and therefore short interaction timescales, such that the cluster is unaffected by the interaction since a short $dt$ minimizes $I_{tide}$. This finding indicates that there exists a minimum interaction timescale, below which clusters are unaffected by close encounters with sub-halos. In fact a minimum interaction timescale should exist for all forms of perturbing substructure (e.g. giant molecular clouds), but will likely depend on the mass and density distribution of the perturbers. It also leads us to conclude that the degree of fluctuation in the tidal tensor alone cannot be used to constrain $T_{diss}/T_{diss,0}$, with the encounter timescale of each tidal interaction and the escape timescale of individual stars being important parameters that need to be taken into consideration as well \citep{spitzer58, webb18b}. 

\subsection{The Effect of Low-Mass Sub-halos on the Tidal Tensor and Cluster Evolution}\label{sec:mass_function}

\citet{penarrubia18} predicts that the degree of fluctuation in the tidal field experienced by a cluster will depend on the allowed range of sub-halo masses. More specifically \citet{penarrubia18} finds that extending the sub-halo mass function to lower and lower masses will increase the degree of fluctuation in the tidal tensor, since the frequency of sub-halo interactions will increase. In fact, \citet{penarrubia18} suggests that low-mass sub-halos may even be the dominant source of fluctuation in the tidal tensor. However it is not necessarily the case that these fluctuations will accelerate cluster disruption \citep{penarrubia19}.

To test how strongly low-mass sub-halos affect cluster evolution, we make use of simulations of clusters in galaxy models where the minimum sub-halo mass has been increased from $10^5$ to $10^6 {\rm M}_\odot$. We find that, for a given substructure fraction and orbital distance, the amount of mass lost by model clusters does not depend on whether or not sub-halos with masses less than $10^6 {\rm M}_\odot$ are included in the galaxy model. Hence cluster evolution is not sensitive to the lower limit of the sub-halo mass function as sub-halos with masses between $10^5 {\rm M}_\odot$ and $10^6 {\rm M}_\odot$ contribute very little to the mass-loss rate experienced by each cluster.

Given the minimal effect that low-mass sub-halos have on cluster evolution, we next determine how each mass range contributes to the tidal tensor. More specifically, we separate the dark matter substructure component of the $\rm{f}_{\mathrm{\rm sub}}=1\%$ galaxy model tidal tensor with point mass sub-halos by sub-halo mass. 

\begin{figure*}
\centering
\includegraphics[width=\textwidth]{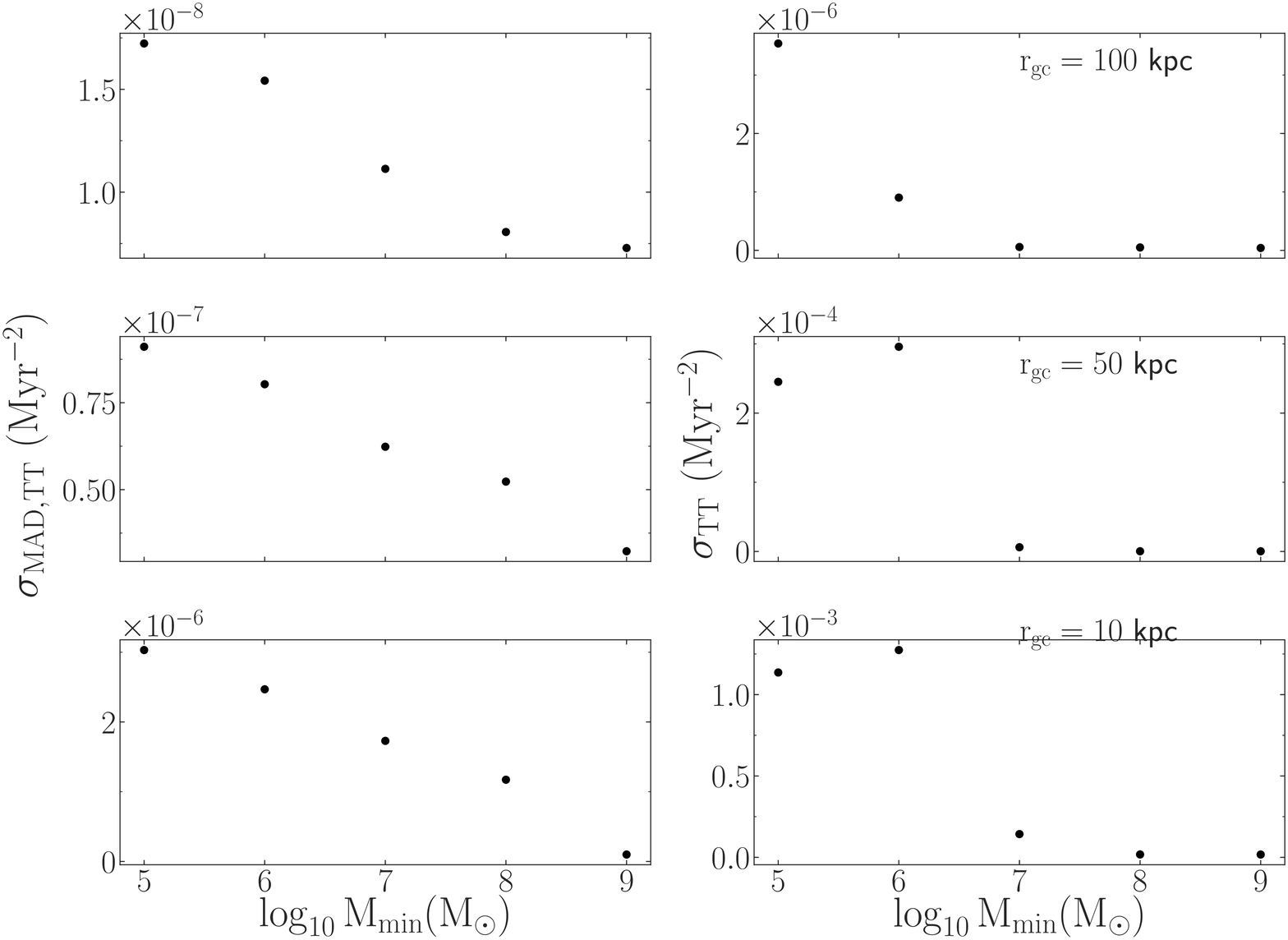}
\caption{Left Panel: Mean absolute deviation of the sum of the squares of the tidal tensor's eigenvalues due to point-mass dark matter substructure as a function of minimum sub-halo mass at galactocentric distances of 10 kpc (bottom), 50 kpc (middle), and 100 kpc (top). Right Panel: Same as the left panel but for the standard deviation of the sum of the squares of the tidal tensor's eigenvalues due to dark matter substructure.}
  \label{fig:sigtt_mad}
\end{figure*}

Figure \ref{fig:sigtt_mad} illustrates that, in agreement with \citet{penarrubia18}, lower mass sub-halos cause a higher degree of fluctuation in the tidal tensor than higher mass sub-halos for the sub-halo mass-function considered here. For a given $\rm{f}_{\mathrm{\rm sub}}$, decreasing the lower limit of the mass function increases the total number of sub-halos in a galaxy. More sub-halos increases the amount of tidal heating, which in Figure \ref{fig:sigtt_mad} is shown by the degree in which $\sigma_{\rm MAD,TT}$ increases as the minimum sub-halo masses decreases. Similarly, more sub-halos also results in the number of direct encounters between sub-halos and clusters, which in turn causes $\sigma_{\rm TT}$ to increase as the minimum sub-halo mass decreases. It is interesting to note that there is significantly more scatter in the relationship between $\sigma_{\rm TT}$ and minimum sub-halo mass than in $\sigma_{\rm MAD,TT}$. The decrease in $\sigma_{\rm TT}$ with minimum sub-halo mass is also much steeper than it is for $\sigma_{\rm MAD,TT}$. Given that $\sigma_{\rm TT}$ serves more as a tracer of tidal shocks, this behaviour indicates that lower mass sub-halos are primarily responsible for tidal shocks and that the degree of fluctuation they impart onto the tidal tensor can vary from one galaxy model realization to the next. More specifically, in Figure \ref{fig:sigtt_mad} it appears that clusters at 10 kpc and 50 kpc undergo at least one close encounter with a $10^6 {\rm M}_\odot$ sub-halo that causes $\sigma_{\rm TT}$ to be higher for $10^6 {\rm M}_\odot$ sub-halos than for $10^5 {\rm M}_\odot$ sub-halo. The increase in $\sigma_{\rm TT}$ is not observed for the cluster at 100 kpc, suggesting the behaviour is driven by small-number statistics and may be relieved by generating multiple realizations of the same galaxy model. Additional simulations are required to test this theory further, especially galaxy models that include sub-halos with larger apocentres.

The mass loss rates of model clusters in galaxies with different minimum sub-halo masses indicate that, despite the increase in $\sigma_{\rm MAD,TT}$ and $\sigma_{\rm TT}$, extending the sub-halo mass function below $10^5 {\rm M}_\odot$ will not increase the effectiveness of sub-halos in disrupting star clusters. Hence tensor fluctuations due to distant low-mass sub-halos are too weak while shocks due to nearby low-mass sub-halos are both too weak and too short to significantly impact a cluster's evolution. Given the results of Figure \ref{fig:sigtt_mad}, allowing for the possibility of the sub-halo mass function extending down to Earth masses or lower will further increase the degree of fluctuation in the local tidal field but will not cause clusters to lose more mass than if the minimum halo was $10^5 {\rm M}_\odot$. Hence when modelling clusters in tidal fields extracted from cosmological simulations, it is more than sufficient if the minimum dark matter sub-halo mass is approximately $10^5 {\rm M}_\odot$.

\subsection{Applicability of Point-Mass Sub-halo Simulations}

With our simulations finding that CDM-like Hernquist sphere sub-halos have no effect on cluster evolution, we use point-mass sub-halos to set the upper limit on how much additional mass loss GCs could possibly experience due to sub-halo interactions. However the results of the point-mass model simulations are not exclusive to point-mass sub-halos only, but any dense sub-halo for which the majority of its mass profile is enclosed within the impact parameter of an interaction with a GC. Therefore it is worthwhile to explore the range of sub-halo densities that could be accurately modelled as point-masses.

\begin{figure}
\centering{}
\includegraphics[width=\columnwidth]{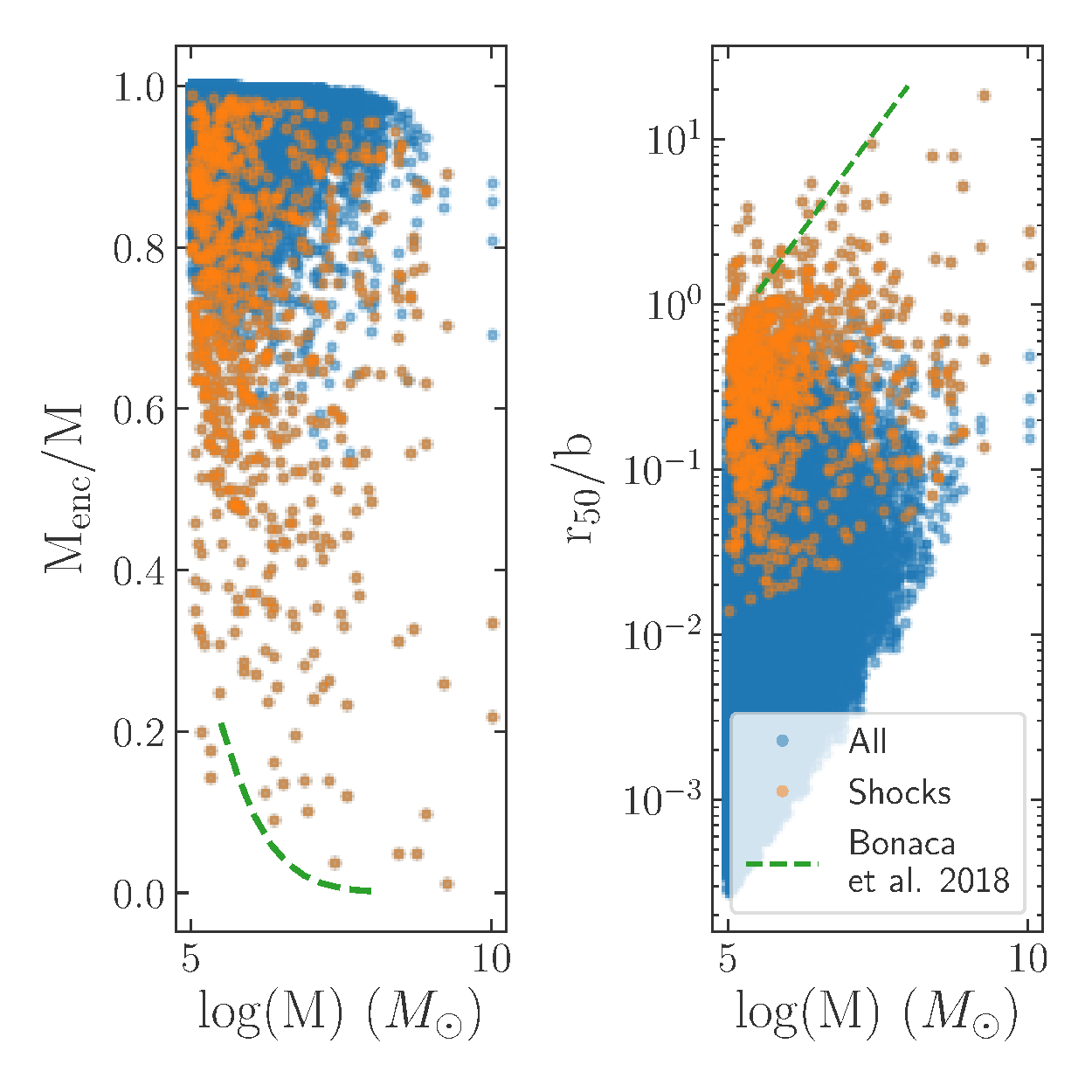}
\caption{Ratio of the mass enclosed within each sub-halo's minimum impact parameter to its total mass (left panel) and the ratio of each sub-halo's half-mass radius to its minimum impact parameter with clusters at 10 kpc, 50 kpc, and 100 kpc over 10 Gyr (blue). Interactions which result in the tidal tensor fluctuating by more than $20\%$ of its amplitude are shown in orange. For comparison purposes, the range of sub-halo and interaction properties that \citet{bonaca18} estimates could be responsible for the GD-1 gap and spur are shown in green, assuming a 1/2 power-law mass-size relation.}
  \label{fig:ttmenc}
\end{figure}

First, we determine the minimum impact parameter that each sub-halo has with the test model clusters on circular orbits at 10 kpc, 50 kpc, and 100 kpc. We then determine what fraction of a CDM-like Hernquist sphere sub-halo's mass is interior to the minimum impact parameter, illustrated as a function of total sub-halo mass in the left panel of Figure \ref{fig:ttmenc} for the $\rm{f}_{\rm sub}=3 \%$ galaxy model; this mass ratio is effectively the ratio of the respective forces acting on the cluster. The right panel of Figure \ref{fig:ttmenc} illustrates the ratio between the radius that contains $50\%$ of the sub-halo's mass and the minimum impact parameter, which serves as an indication of how much smaller a sub-halo of equal mass would have to be in order to be viewed as a point-mass by cluster stars. For illustrative purposes, sub-halos which cause a strong tidal shock (which we loosely define as a fluctuation in the tidal tensor greater than $20\%$ of its amplitude) are shown in orange. We also show the range of sub-halo masses that \citet{bonaca18} estimate a perturbing sub-halo must have, along with an impact parameter of 50 pc, in order to have created a gap and spur in the GD-1 stream. While \citet{bonaca18} estimates the sub-halo must have had a scale radius of 20 pc, we initially assume the sub-halos follow a 1/2 power-law mass-size relation ($\rm{r_s}=1.05$ kpc $(\rm{M}/10^8 {\rm M}_\odot)^{\frac{1}{2}}$) to best compare with the models.

Figure \ref{fig:ttmenc} demonstrates that most of the CDM-like Hernquist sphere sub-halos can actually be assumed to be point masses as they do not pass close enough to a model cluster for $\rm M/M_{enc}$ to be significantly less than unity. However there are several close encounters between sub-halos and clusters where forces from point-mass sub-halos will be an order of magnitude stronger than CDM-like Hernquist sphere sub-halos. With respect to cluster sizes, the majority of sub-halos have half-mass radii significantly smaller than the minimum impact parameter, confirming that most sub-halos can be treated as point-masses. For sub-halos causing strong tidal shocks, point-mass sub-halos can serve as a proxy for Hernquist sphere sub-halos that are one to at most two orders of magnitude smaller than our assumed mass-size relation yields.

Recent work by \citet{bonaca18} considered on the properties of a possible dark matter sub-halo that could produce a gap and spur in the GD-1 is in support of sub-halos being dense enough that interactions can be modelled using point-mass sub-halos. Figure \ref{fig:ttmenc} indicates that following our mass-size relation would result in less than $20\%$ of the perturbing sub-halos mass being within the interactions impact parameter. However \citet{bonaca18} finds these sub-halos must have a significantly smaller scale radius (20 pc) such that half of the sub-halo's mass is within the impact parameter. Therefore, the proposed sub-halos must be between 3 and 50 times smaller (one to five orders of magnitude denser) in order for $\rm M_{enc}/M$ to equal 0.5 and $\rm r_{50}/b$ to be less than 1. Assuming the gap and spur in GD-1 are in fact caused by a sub-halo interaction, a compact Hernquist sphere sub-halo that was half as massive and two orders of magnitude smaller (such that the entirety of its mass is within the impact parameter) would have a similar effect and also be accurately modelled as a point-mass sub-halo. Hence, in addition to the point-mass galaxy models placing upper limits on how strongly clusters can be affected by dark matter substructure, they also serve as a proxy for how different mass-size relations will alter cluster evolution.

Overall, for the substructure fractions considered here ($\rm{f}_{\rm sub} < 10\%$), it appears that an indirect detection of dark matter substructure using globular clusters is only possible if sub-halos are \textit{very} compact. Furthermore, a cluster must undergo a close encounter with a relatively massive ($\rm{M} >10^6 {\rm M}_\odot$) sub-halo which, given a power-law mass-function of $-1.9$, will only occur on the order of unity per Hubble time. For clusters that have been found to have lost more mass than their current orbit and structure indicate, as found by \citep{webb15}, the additional mass loss could be attributed to interactions with compact sub-halos. Higher substructure fractions would lead to more frequent sub-halo interactions, which may relax our finding that sub-halos must be massive and compact to significantly alter cluster evolution. In the case that sub-halos are extended, which is more in line with cosmological simulations, tidal shocks can still serve as an additional source of mass loss but the perturbing source would then have to be baryonic \citep[e.g.][]{gieles06,elmegreen10,kruijssen15,banik18b}.

Extending the suite of simulations to higher mass clusters, in line with observed globular clusters, may also increase the ability of sub-halos to affect star cluster evolution. For example, clusters that are $10\times$ more massive than the model clusters considered here will be over $2 \times$ larger (assuming equal densities). Hence not only will two-body relaxation play a lesser role in the more massive cluster's evolution, but it will undergo closer and more frequent encounters with dark matter sub-halos (increasing both $\sigma_{\rm MAD,TT}$ and $\sigma_{\rm TT}$). Similarly, extended clusters like Pal 4 and Pal 14 that simulations have been unable to reproduce \citep{zonoozi11, zonoozi14} and ultra-faint objects \citep{contenta17} (all of which are typically low in mass and found in the outer halo where $\rm{f}_{\rm sub}$ is expected to be higher) will be more strongly affected by tidal heating than the model clusters considered here. Combined with the fact that our simulations likely underestimate the effect that sub-halos have on cluster evolution, due to the tidal approximation underestimating the force acting on outer region stars by passing sub-halos, it is critical that future studies are able to directly model the effects of dark matter sub-halos on the evolution of massive star clusters and extended star clusters.

\section{Conclusion}\label{s_conclusion}

We have modelled the evolution of tidally filling 30,000 ${\rm M}_\odot$ star clusters orbiting at 10 kpc, 50 kpc and 100 kpc in galaxies with varying fractions of dark matter substructure and sub-halo properties. In our galaxy models, the orbits of sub-halos with masses between $10^5$ are $10^{11} {\rm M}_\odot$ are solved and the tidal tensor experienced by each cluster is calculated over 10 Gyr of evolution. The time resolution of the tensor calculations are set by the shortest interaction timescale found in the simulation, ensuring that we fully resolve each close encounter between a sub-halo and a cluster. 

Overall, our simulations demonstrate that orbiting in a galaxy with a non-zero fraction of mass in the form of dark matter substructure leads to significant fluctuations in the tidal tensor experienced by a cluster, which results in clusters being tidally heated due to long-duration distant sub-halo encounters and short-duration close encounters (tidal shocks). However, modelling dark matter sub-halos as Hernquist spheres with a power-law mass-size relation that has a slope of 0.5 has a minimal effect on star cluster evolution, as the corresponding tidal fluctuations are either too small or do not last long enough for stars to be energized to the point of escaping the cluster. Treating sub-halos as point sources on the other hand increases the strength of individual interactions, which increases fluctuations in the tidal tensor experienced by a cluster such that heating results in an additional source of mass loss in clusters over the course of their lifetime. A similar result is expected for Hernquist sphere sub-halos that are two to three orders of magnitude smaller than those considered here, in agreement with \citet{penarrubia18}. Analyzing the dissolution times of clusters interacting with point-mass sub-halos, relative to the dissolution time of initially identical clusters in smooth dark matter potentials, suggests that close encounters are the main mechanism behind unbinding stars from a cluster over long-duration distant encounters. 

For a given $\rm{f}_{\mathrm{\rm sub}}$, we also find that the mass lost by clusters due to sub-halo interactions is not sensitive to the lower-limit of the sub-halo mass limit. Increasing the lower limit of the sub-halo mass function from $10^5$ to $10^{6} {\rm M}_\odot$ results in clusters losing the same amount of mass despite undergoing fewer sub-halo interactions. Hence the mass loss experienced by clusters interacting with point-mass sub-halos can be attributed to high-mass sub-halos becoming denser. Furthermore, in agreement with \citep{penarrubia19}, we find that the increased degree of fluctuation in the tidal tensor caused by including low-mass sub-halos does not result in an increase in cluster mass loss rates. The fluctuations are either too weak, too short, or both. In fact, it is likely that extending the sub-halo mass function below $10^{5} {\rm M}_\odot$ will have no effect on cluster evolution. Hence when modelling star clusters in tidal fields extracted from cosmological simulations, it is not necessary to include dark matter sub-halos with masses less than $10^{5} {\rm M}_\odot$ for the substructure fractions considered here.

Modelling interactions with massive point-mass sub-halos also serves as a tracer for interactions with massive Hernquist sphere sub-halos that are so compact that their entire mass profile is within the interaction's impact parameter. We find that such interactions can lead to smaller cluster sizes. The evolution of the density profiles of the clusters, however, shows no clear dependence on $\rm{f}_{\mathrm{\rm sub}}$. A weak kinematic signature does exist in both the velocity dispersion and anisotropy profiles of clusters subject to massive point mass sub-halo interactions. For clusters that have recently undergone a tidal shock, short lived spikes in $\sigma_{\rm v}$ are observed in the outskirts of cluster. While this signature typically disappears within a crossing time, over the course of a cluster's lifetime repeated shocks result in a small increase in $\sigma_{\rm v}$ in the outermost regions of a cluster relative to the no-substructure case. Furthermore, for point mass sub-halos, the anisotropy profile of the cluster also retains knowledge of the fact that the cluster has undergone tidal heating and repeated shocks when the external tidal field is weak, as sub-halo interactions are only able to remove stars on eccentric orbits. For clusters that orbit in strong tidal fields, however, the signature does not exist as the strong tidal field can remove stars over a wide range of orbits. Hence an observed increase in $\sigma_{\rm v}$ and decrease in $\beta$ in the outskirts of distant clusters serves as a potential signature that a cluster has been interacting with massive and compact substructure over the course of its lifetime. Recent work on sub-halo interactions with stellar streams by \citet{bonaca18} indicates that sub-halos may come close to reaching the densities and masses required to affect the observed properties of clusters. 

With cosmological simulations predicting that $\rm{f}_{\mathrm{\rm sub}}$ is over an order of magnitude higher in the outskirts of the Milky Way where the tidal field is weak, globular clusters at large galactocentric distances may prove to be the key to unlocking the properties of dark matter. Unfortunately, only inner region clusters have had their anisotropy profile accurately measured to date \citep[e.g.][]{watkins15, bellini17, milone18}. Furthermore, a decrease in $\beta$ with clustercentric distance is also a signature of cluster rotation \citep{vesperini14}, hence rotation must be accounted for before attributing tangential anisotropy in the outskirts of a cluster to dark matter substructure interactions. Both current (Gaia \citep{gaia16} and the High-resolution Space Telescope Proper Motion Collaboration \citep{vandermarel14}) and planned future missions (CASTOR \citep{cote12}, WFIRST \citep{spergel15}, GIRMOS \citep{sivanandam18}) offer the capability of measuring rotation and orbital anisotropy in distant clusters and potentially confirming the existence of dark matter substructure in the Milky Way if massive sub-halos are sufficiently dense. Otherwise a null detection can still help constrain the properties of dark matter as it will place a lower limit on sub-halo scale radii. It should also be noted that unless the cluster has very recently undergone a tidal shock, it may not be possible for these missions to attribute small changes in the velocity and anisotropy profiles of clusters to dark matter substructure only.

It will also be important to include a wide range of initial cluster properties in future work, as a cluster's mass and density play important roles in how it will respond to dark matter sub-halo interactions. Including a stellar mass spectrum and stellar evolution will also be necessary, as both lead to cluster expansion. Additional studies aimed at quantifying the effect that sub-halos have on globular clusters will allow us to probe several properties of dark matter, including the present day $\rm{f}_{\mathrm{\rm sub}}$, the sub-halo mass function, the sub-halo mass-size relation, and help constrain the nature of dark matter itself. It may also prove true that sub-halos are partially responsible for shaping the distribution of cluster masses and sizes in a galaxy. Continuing to model clusters in time dependent and realistic environments will help push the field of globular cluster studies towards truly being able to use clusters to constrain both their own formation and evolution history and that of their host galaxy.

\section*{Acknowledgements}
We would like to thank the referee for feedback and suggestions on how to improve the submitted version of the manuscript. JW acknowledges financial support through a Natural Sciences and Engineering Research Council of Canada (NSERC) Postdoctoral Fellowship. JW and JB also acknowledge additional financial support from NSERC (funding reference number RGPIN-2015-05235) and an Ontario Early Researcher Award (ER16-12-061). JW also would like to thank Marta Reina-Campos, Diederik Kruijssen, and Jorge Pe\~{n}arrubia for valuable discussions and feedback. MG acknowledges support from the European Research Council (ERC StG-335936) and the Royal Society (University Research Fellowship). MG acknowledges Denis Erkal and Justin Read for several interesting discussions. This work was made possible in part by the facilities of the Shared Hierarchical Academic Research Computing Network (SHARCNET:www.sharcnet.ca) and Compute/Calcul Canada, in part by Lilly Endowment, Inc., through its support for the Indiana University Pervasive Technology Institute, and in part by the Indiana METACyt Initiative. The Indiana METACyt Initiative at IU is also supported in part by Lilly Endowment, Inc.

\bsp

\label{lastpage}

\end{document}